\documentclass[12pt,preprint]{aastex}

\shorttitle{A Radio Spectral Line Study of the 2-Jy IRAS-NVSS Sample}
\shortauthors{Fernandez et al.}

\newcommand{\HI}{{H~{\footnotesize I }}}

\begin{document}

\title{A Radio Spectral Line Study of the 2-Jy IRAS-NVSS Sample: Part I}

\author{Maria Ximena Fernandez}
\affil{Department of Astronomy, Columbia University, 550 West 120th Street, New York, NY 10027}
\email{ximena@astro.columbia.edu}

\author{Emmanuel Momjian}
\affil{National Radio Astronomy Observatory, P. O. Box O, Socorro, NM, 87801}
\email{emomjian@nrao.edu}

\author{Christopher J. Salter}
\affil{NAIC, Arecibo Observatory, HC 3, Box 53995, Arecibo, PR 00612}
\email{csalter@naic.edu}

\author{Tapasi Ghosh}
\affil{NAIC, Arecibo Observatory, HC 3, Box 53995, Arecibo, PR 00612}
\email{tghosh@naic.edu}

\begin{abstract}
We present results from an on-going survey for the \HI 21 cm line and the OH
18~cm lines in IR galaxies with the Arecibo 305~m Radio Telescope.
The observations of 85 galaxies extracted from the 2 Jy IRAS-NVSS sample
in the R.A.~(B1950) range 20$^{\rm h}$--00$^{\rm h}$ are reported in
this paper.  We detected the \HI 21~cm line in 82 of these galaxies, with 18
being new detections, and the OH 18~cm lines in 7 galaxies, with 4 being new
detections. In some cases, the \HI spectra show the classic double-horned
or single-peaked emission profiles. However, the majority exhibit
distorted  \HI spectral features indicating that the galaxies are in
interacting and/or merging systems. From these \HI and OH observations,
various properties of the sample are derived and reported.
\end{abstract}


\section{Introduction}
At bolometric luminosities above $10^{11} L_ {\odot}$, infrared (IR)
galaxies become the dominant population of extragalactic objects in the
local universe ($z \leq 0.3$). These galaxies are subdivided into three
categories: luminous (LIRGs, $L_{\rm IR} > 10^{11} L_ {\odot}$),
ultraluminous (ULIRGs, $L_{\rm IR} >10^{12} L_ {\odot}$), and
hyperluminous (HyLIRGs, $L_{\rm IR} > 10^{13} L_ {\odot}$; \citealp{SM96}).
Even though these IR galaxies are relatively rare, comprising less
than 6\% of the total IR energy density in the local Universe \citep{SN91}, some
studies suggest that the majority of galaxies with $L_{\rm B} > 10^{11}
L_ {\odot}$  go through a stage of intense IR emission \citep{SOI87}.

Most IR galaxies with $L_{\rm IR} < 10^{11} L_{\odot}$ are single, gas-rich spirals, and
their IR emission can be accounted for by star formation.
In the luminosity range $10^{11} < L_{\rm IR} < 10^{12} L_ {\odot}$,
most of the galaxies are interacting/merging systems with enormous quantities of molecular
gas ($\sim 10^{10}~M_ {\odot}$). At the lower end of this range, the bulk of the IR
luminosity is due to warm dust grains heated by a nuclear
starburst, while active galactic nuclei (AGN) become increasingly
important at higher luminosities. Galaxies with $L_{\rm IR} >10^{12}
L_ {\odot}$ are believed to be advanced mergers powered by a
combination of starburst and AGN \citep{SM96}.

Previous \HI observations of IR galaxies have revealed very broad
absorption lines in ULIRGs, indicating rotation plus large amounts
of turbulent gas \citep{MIR82}. High angular resolution Very Large Array (VLA)
and Very Long Baseline Interferometry (VLBI)
observations show that these galaxies have the absorbing \HI situated
in the inner few hundred parsecs along the line of sight to the
nuclear continuum sources \citep{BGSM87,MOM03}.

Several OH 18~cm absorption and megamaser (hereafter OHM)
emission surveys of ULIRGs have also been published 
\citep{B89,DG00,DG01,DG02}. \citet{B89} concluded that the OHM
emission usually occurs in galaxies with higher far-IR (FIR)
luminosities and flatter 100--25~$\mu$m spectra rather than in those
with OH 18~cm absorption features.

Here, we report results from an on-going spectroscopic survey with the
Arecibo Radio Telescope\footnote {The Arecibo Observatory is part of the
National Astronomy and Ionosphere Center, which is operated by Cornell
University under a cooperative agreement with the National Science
Foundation.} targeting the \HI 21~cm and the main and satellite OH 18~cm
lines of 85 IR
galaxies from the 2-Jy IRAS-NVSS sample \citep{YRC01}.  In this paper,
we adopt  $H_0$ = 71 km $\rm s^{-1}$ $\rm Mpc^{-1}$, $\Omega_{M}$ =
0.27, and $\Omega_{\Lambda}$ = 0.73.

\section{The Sample}
The 85 galaxies reported in this paper are a first set of objects
extracted from the 2-Jy IRAS-NVSS sample \citep{YRC01}, which consists
of 1809 IRAS sources that have $S_{60~\mu m} \geq  2$~Jy with 1.4~GHz radio
counterparts from the NRAO-VLA Sky Survey (NVSS; \citealp{CON98}). The
selection criteria for our full sample are: (1) they lie within the
area of sky accessible to the Arecibo telescope  (i.e. $-1^{\circ}
<$ declination $< 38^{\circ}$), (2) have $L_{\rm FIR} \geq 7 \times
10^{9}~L_{\odot}$, and (3) have heliocentric velocities between 0
and 50,000 km~s$^{-1}$. The number of galaxies that meet these criteria
is 582. The present study includes 85 of these galaxies within the
R.A. (B1950) range 20$^{\rm h}$--00$^{\rm h}$.

Table~1 presents the 85 galaxies reported in the paper. Column (1) gives
the IRAS names of the galaxies, and column (2) lists other common designations,
where applicable. Column (3) and (4) are the Right Ascensions
and Declinations actually observed in J2000 coordinates. Column (5) lists
the optical redshifts retrieved from the NASA/IPAC Extragalactic Database
(NED). Column (6) is the 60~${\mu}\rm m$ flux densities from the IRAS
Faint Source Catalog (FSC), where available. Galaxies marked with an
asterisk do not have ${60~{\mu}\rm m}$ flux density measurements in
the FSC, and for these the IRAS Point Source Catalog (PSC) values are
given instead. Column (7) is the 1.4~GHz flux densities from the NVSS
survey. Column (8) provides the morphologies listed in NED and in the
Hyperleda database \citep{PAT03}.

\section{Observations and Data Reduction}

The 85 galaxies were observed with the 305~m Arecibo radio telescope
between July and November 2004 using the L-band wide receiver. The low
frequency cut-off of the receiver is 1100~MHz, well below the
heliocentric velocity limit of 50,000 km $\rm s^{-1}$ for our sample.
The four independent boards of the Arecibo Interim Correlator were utilized to
simultaneously observe the following redshifted transitions: the \HI
21~cm line (1420.40575~MHz), both of the OH 18~cm main lines (1665.4018 and
1667.359~MHz), and the two OH 18~cm satellite lines at 1612.231 and 
1720.530~MHz.  A bandwidth of 12.5~MHz (dual
polarization) was used for each correlator board with 1024 spectral
channels per polarization. The three boards that targeted the \HI
and the OH satellite lines were centered at the frequencies of the
respective redshifted transitions, while the central frequency of the
board targeting the two OH 18~cm main lines was set to the redshifted value of
the 1667.359~MHz transition.
We used the total-power, position-switching (ON-OFF),
observational mode for all the objects in our sample. However, the spectra of some
sources showed significant baseline ripples. For these sources, the
Double Position Switching (DPS; \citealp{ghos02}) observing technique,
which requires observing a nearby bandpass calibrator,
was utilized in order to minimize the baseline ripples in their spectra.

The data were reduced using the standard Arecibo Observatory (AO)~IDL
routines and special-purpose codes developed by AO staff. The spectra of
the sources observed with the ON-OFF mode were converted to units of flux
density using the noise-diode calibration and the standard gain curves. For sources
observed in DPS mode, the flux densities of the bandpass calibrators were
used to convert the ratio spectra to Janskys. Hanning smoothing was
applied to the raw spectra, and after averaging the two polarizations,
a 9-channel boxcar smoothing was applied to further improve the signal-to-noise
ratio. The resulting spectral resolution was 112~kHz.
Polynomial baselines were fitted to the resulting spectra and
subtracted from them to produce the final spectra used to derive the velocity,
velocity widths, the integrated intensities of the lines, and the rms noises on the
spectra for each object.

\section{Results}

\subsection{\HI 21 cm Line Spectra}

We detected 21 cm \HI in emission
and/or absorption in 82 of the 85 galaxies;
18 of these had not been previously reported in the
literature. There were 7 sources with both absorption and emission
features in their spectra, and 3 with pure absorption. IRAS
21396+3623, one of the three non-detections, was observed at an erroneous
frequency because its redshift was incorrectly listed in the 2-Jy IRAS
NVSS Sample. This source has not been included in the analysis of the
data.

Figure~1 shows the spectra of the 18 new \HI 21~cm detections, while Figure~2
shows our spectra for those sources with previously published \HI 21~cm detections.
We note that Figures~1 and 2 present the spectra before subtraction
of the fitted polynomials. These hence demonstrate the quality of the raw baselines,
and reveal any continuum emission associated with the source.

Some of the 21~cm \HI emission spectra in our sample show single peak or
double horn profiles, both characteristic of non-interacting galaxies.
However, the majority of the galaxies observed have distorted
spectra as expected for interacting/merging systems. The
galaxies in such systems have higher IR luminosities than do
non-interacting spirals.

Table~2 summarizes the physical properties derived from the \HI emission
spectra of the galaxies in our sample, including non-detections (marked
by an asterisk). The values in this table are derived using the
baseline-subtracted spectra.
Sources marked by $^\diamond$ are those observed
in DPS mode. Seven sources (indicated by $^\dag$ in Table~2) exhibit
both emission and absorption features in their spectra. For these, the
table contains information from the emission part of the spectra, and
the derived flux density integrals and neutral hydrogen masses should
be considered to be lower limits because some of the emission might be
masked by the absorption features. The absorption components are presented
separately later in this section. Column (1) in Table~2 lists the IRAS
names of the galaxies. Column (2) gives the total on-source integration
times. Column (3) lists the rms noise levels. Column (4) lists the heliocentric velocities at
which the \HI emission lines are centered. For non-detections, the values
correspond to the optical velocities retrieved from NED. Column (5) is
the full velocity widths at half maximum (FWHM) for the observed
emission lines. Column (6) is the flux density integrals ($\int{S dv}$) of the \HI lines.
For non-detections, the
values are 3$\sigma$ upper limits derived assuming $\Delta{V_{50}} =
400$~km~s$^{-1}$. Column (7) lists the luminosity distances ($D_{\rm L}$) of the galaxies
derived using their respective velocities (column 4).
Column (8) gives the logarithms of the FIR luminosities calculated using
the equation (e.g. \citealp{SM96}):
\begin{equation}
L_{\rm FIR} (L_ {\odot}) = 3.96 \times 10^{5}~D_{\rm L}^{2}~(2.58f_{60}+f_{100})
\end{equation}
where $D_{\rm L}$ is in Mpc, and $f_{60}$ and $f_{100}$ are respectively
the 60 and 100~$\mu$m flux densities in Jy. Column (9) is the logarithms of the total
IR luminosities derived using the equation (e.g. \citealp{SM96}):
\begin{equation}
L_{\rm IR} (L_ {\odot})=5.67 \times 10^{5}~D_{\rm L}^{2}~(13.48f_{12}+5.16f_{25}+2.58f_{60}+f_{100}).
 \end{equation}
Column (10)
lists the logarithms of the total neutral hydrogen mass values derived using
the expression \citep{Rob75} :
\begin{equation}
M_{\rm HI} (M_ {\odot})= 2.36 \times 10^{5}~D_{\rm L}^{2} \int{S dv}
\end{equation}
where the $\int{S dv}$ is the flux density integral in Jy~km~s$^{-1}$.

Table~3 lists the parameters of the \HI absorbers. The properties of the \HI absorption lines
are derived from spectra where the flux
densities were scaled into optical depths ($\tau$) and plotted against heliocentric
velocity. The values in this table are also obtained using the
baseline-subtracted spectra.
Column (1) lists the IRAS names of the galaxies. Column (2) lists their luminosity distances.
These distances are derived using the velocity of the \HI emission line for sources that
exhibit both \HI emission and absorption (from Table~2),
and the optical redshift values (from Table~1) for sources that exhibit \HI absorption only.
Columns (3) and (4) give the logarithms of the FIR and the total IR luminosities
calculated using equations~1 and 2, respectively. Column (5) is the total
on-source integration times. Columns (6) lists the rms noise levels.
Column (7) lists the
heliocentric velocities at the peak of the absorption feature in each source.  Column
(8) is the FWHM velocity values of the overall observed absorption profile. Column (9)
lists the peak optical depths. Column (10) gives the \HI column densities
divided by the spin temperature which are derived assuming a covering
factor of unity and using the equation (e.g. \citealp{Roh86}):
\begin{equation}
N({\rm HI})/ T_{\rm s}~(\rm cm^{-2}~K^{-1}) = 1.823 \times 10^{18} \int{\tau dv}.
\end{equation}

\subsection{OH 18 cm Line Spectra}
We detected OH 18~cm main lines in 7 galaxies; 3 in emission (OHMs)
and 4 in absorption. Of these, 1 OHM and 3 OH absorbers are new
detections (Figure~3). For these spectra, the 1667.359 MHz line was used
for deriving the velocity scale. There were no detections of the OH 18~cm
satellite lines at 1612 and 1720 MHz in any of the galaxies of our sample.
Figure~4 shows the spectra of the 3 sources with previously known OH 18~cm main lines.
As for the \HI 21 cm spectra in Figures 1 \& 2, Figures~3 \& 4 present the final
spectra, but without the subtraction of fitted polynomials.

Table~4 summarizes the parameters derived from the OHM detections.
The values in this table are derived using the
baseline-subtracted spectra.
Column (1) lists the IRAS names of the galaxies. Column (2) is the total on-source
integration times. Column (3) gives the rms noise levels, and column (4)
lists the heliocentric velocities at the center of the 1667.359~MHz line
emission. Columns (5) and (8) are the peak flux densities of the 1667 and 1665~MHz
lines respectively, columns (6) and (9) are their full velocity widths
at half maximum, and columns (7) and (10) list their integrated flux densities.
Column (11) lists the hyperfine ratios
obtained by dividing the integrated flux density of the 1667~MHz line by
that at 1665~MHz for each source. In thermodynamic equilibrium conditions, this ratio
would be $R_H=1.8$, and it increases as the saturation of a masing region
increases. Column (12) lists the logarithms of the predicted OH luminosities (in units
of ${L_{\odot}}$) which have been calculated using the following equation from \citet{KAN96}:
\begin{equation}
{\rm log} {L_{\rm OH}^{\rm pred}}=1.38~{\rm log} {L_{\rm FIR}}-14.02,
\end{equation}
while column (13) gives the logarithms of the measured isotropic OH line
luminosities, which represents the combined integrated flux densities of the two OH
18~cm main lines.
Columns (14) and (15) are the rms
noise values for the 1612 and 1720~MHz transitions. Blank fields in these two
columns indicate that the rms noise values could not be estimated because of severe radio
frequency interference (RFI).

Table~5 lists the parameters derived from the OH absorption lines.
The values in this table are derived using the
baseline-subtracted spectra.
Column (1) is the IRAS names of the galaxies.  Column (2) is the total on-source
integration times. Column (3) gives the rms noise levels, and column
(4) lists the heliocentric velocities at the center of the 1667.359~MHz
absorption lines.
Columns (5) and (9) are the full velocity widths at
half maximum of the 1667 and 1665~MHz lines respectively, columns (6)
and (10) are their peak optical depths,
and
columns (7) and (11) are their integrated optical depths.
Column (8) is the column densities of the 1667~MHz
lines divided by the excitation temperature using the following equation from \citet{Tur73}
and assuming a covering factor of unity:
\begin{equation}
N_{\rm OH}/\rm T_{ex} ({\rm cm^{-2}~K^{-1}}) = 2.35 \times 10^{14}  \int{\tau dv}
\end{equation}
where $\int{\tau dv}$ is the integrated optical depth of the 1667~MHz
absorption feature given in column (7).  Column (12) lists the hyperfine
ratios, obtained by dividing the integrated optical depths of the two
main lines.
Columns (13) and (14) are the rms
noise values for the 1612 and 1720~MHz transitions. Blank fields in
these two columns indicate that the rms noise values could not be estimated
because of severe RFI.

Table~6 lists relevant data for all sources with no detections in the
OH 18~cm mainlines. Column (1) is the IRAS names of the galaxies. Column
(2) lists the rms noise values of the OH 18~cm main line non-detections.
Column (3) is the logarithms of predicted OH luminosities from
equation~5, and column (4) is the logarithms of the maximum OH luminosities
for the 1667~MHz main line determined using the equation (from \citealp{DG00}):
\begin{equation}
L_{\rm OH}^{\rm max}=4\pi~ D_L^2 ~1.5\sigma ~\frac{\Delta V}{c}~\frac{\nu_o}{(1+z)}
\end{equation}
where $D_{\rm L}$ is the luminosity distance, $\sigma$ is the rms
noise value listed in column (2), $\Delta V= 150~ \rm km~s^{-1}$, $c$
is the speed of light, $\nu_o$ is the rest frequency of the 1667~MHz
transition, and $z$ is the redshift. Columns (5) and (6) are the rms
noise values of the 1612 and 1720~MHz transitions.  Blank fields in
this table indicate that the rms noise values could not be estimated because
of severe RFI.

\section{Analysis and Discussion}
Here we report preliminary statistical analysis for the observed
sample.  More detailed analysis and conclusions will be presented
after completion of the full survey with all 582 galaxies.

Figure~5~({\it top}) shows the well established radio--FIR
correlation through a logarithmic plot of the 1.4~GHz continuum
luminosities versus the FIR luminosities for the galaxies in our
sample.  The derived correlation coefficient is 88\%. Figure~5~({\it
bottom}) shows a logarithmic plot of the 1.4~GHz continuum
luminosities versus the total IR luminosities. The correlation
coefficient here is 89\%.

This remarkably tight linear correlation between the total radio
continuum emission and the IR (or FIR) luminosities is well known for
``normal'' galaxies where the main energy source is not due to a
supermassive black hole \citep{CON92}. The most obvious interpretation
of this correlation is the presence of massive stars that provide both
relativistic particles via subsequent supernova events, and heat the
interstellar dust which radiates at IR (or FIR) wavelengths
\citep{HSR85,WK88,CON92}.

Figure~6 shows logarithmic plots of the \HI mass versus the FIR ({\it
top}) and IR ({\it bottom}) luminosities for the observed sample.
Both plots show extremely weak correlations with coefficients of 42\%
in each. Figure~7 shows a logarithmic plot of the \HI mass versus
the 1.4~GHz radio luminosity. The correlation coefficient here is 53\%.
These plots suggest that the total neutral gas content and star
formation activity traced through the radio luminosities or the IR
luminosities are only weakly correlated for this sample. This is
consistent with the scenario that atomic gas has first to be converted
into molecular gas to form stars, and that the molecular gas content
itself correlates well with star formation \citep{WB02}.
In Table 7, we present the mean and median \HI mass values of galaxies with
\HI 21 cm emission as a function of total-IR luminosity bins. The numbers reflect a
general trend of higher \HI mass values at higher IR luminosities, consistent
with the weak correlation seen in Figure~6-{\it bottom}.
We utilize the values presented in this table in our notes on individual objects
in \S 6.

In our observed sample, several galaxies show either \HI absorption or both
\HI emission and absorption. Binning the sample in $L_{\rm IR}$ (Table~8a)
reveals that sources with higher IR luminosities have the greater
likelihood of showing \HI absorption. For instance, 38.5\% of the
sources with $L_{\rm IR} \geq 10^{11.50}~L_{\odot}$ show \HI
absorption, while only 10.3\% of sources with $10^{11.00}~L_{\odot}
\leq L_{\rm IR} \leq 10^{11.49}~L_{\odot}$, and 6.3\% of sources with
$10^{10.50}~L_{\odot} \leq L_{\rm IR} \leq 10^{10.99}~L_{\odot}$ show
\HI absorption. No \HI absorption is seen in sources with $L_{\rm IR}
\leq 10^{10.49}~L_{\odot}$ (see Table 8a). Thus, while we cannot say
anything about traditionally defined ULIRGs ($L_{\rm IR} \geq
10^{12}~L_{\odot}$) because of their rarity in our observed sample,
it appears that galaxies with $L_{\rm IR} \geq 10^{11.50}~L_\odot$ have
a greater likelihood of showing \HI absorption. We now explore whether
this trend could arise due to a selection effect.

While the flux density of \HI emission is proportional to $L_{\rm
HI}/D^2$, where $D$ is the distance to the galaxy, the flux density dip
of an absorption corresponding to a given optical depth is
proportional to the background continuum flux density that is being
absorbed. In Table~8b we present the calculated mean and median
NVSS flux densities for the galaxies in our observed sample as a
function of $L_{\rm IR}$ bins, to investigate whether galaxies having
$L_{\rm IR}\geq 10^{11.50}~L_{\odot}$  preferentially have higher flux
densities. Table~8b shows that the radio flux densities in our sample
do not correlate with IR luminosity.
However, we do find for those galaxies showing \HI absorption in the two highest luminosity bins,
that the mean flux densities are about double the mean values for all galaxies in those bins.
As a second
possible effect, the $L_{\rm IR} \geq 10^{11.50}~L_{\odot}$ absorbers
might have a higher covering factor
than those of lower luminosity
e.g.,
their continuum emission may mostly be in a compact
nucleus
vs.~more extended continuum emission for the others, with the absorption arising principally in the nuclear regions.
However, whether this is the case or not, we do find a difference in the incidence of \HI absorption
between IR galaxies with $L_{\rm IR} \geq 10^{11.50}~L_{\odot}$ and those with lower IR luminosities,
i.e., the galaxies with higher IR luminosities have higher \HI column densities on the lines of sight
to the continuum sources, and/or their continuum emission is confined to more compact regions.

In Table 9, we present the statistics of OH detections (absorption or
emission) as a function of $L_{\rm IR}$ bins. The total number of 
sources per $L_{\rm IR}$ bin in this table excludes galaxies with OH main
line spectra severely affected by RFI. Here we note that all 3 of the OH
emitters have $L_{\rm IR} \geq 10^{11.50}~L_{\odot}$, while the two
detections for $10^{10.50}~L_{\odot} \leq L_{\rm IR} \leq
10^{10.99}~L_{\odot}$ are both absorbers.  However, despite the small
numbers in this subgroup, it is strongly suggestive that 
OH-detected sources (emitters and absorbers) are found primarily in
galaxies with $L_{\rm IR} \geq 10^{11.50}~L_{\odot}$.

\section{Notes on Individual Objects}

{\bf IRAS 20210+1121:}
This galaxy is at a redshift of $z=0.0564$, and is one of two sources with no \HI detection
in our observed sample, with the other source being IRAS~23410+0228 ($z=0.0912$;
see below). Its IR
luminosity is $10^{11.90 \pm 0.09}~L_{\odot}$ and the derived 3$\sigma$ upper limit for its \HI mass,
assuming a 400~km~s$^{-1}$ velocity width, is $10^{9.86}~M_{\odot}$. In the sample presented
in this paper, sources with $L_{\rm IR} > 10^{11.50}~L_{\odot}$ and detected \HI emission 
have a \HI median mass value of $10^{9.86}~M_{\odot}$ and a mean value of $10^{9.99 \pm 0.10}~M_{\odot}$.
This suggests that IRAS~20210+1121 is likely to be hydrogen deficient. However, at IR
luminosities $> 10^{11.50}~L_{\odot}$, there are four sources in our sample with \HI detections
that have \HI masses $<10^{9.86}~M_{\odot}$. Therefore, observations with longer integration time
are needed to establish whether IRAS~20210+1121 is truly hydrogen deficient.

{\bf IRAS 20332+0805:}
Our estimated heliocentric velocity for this galaxy of $7967.3 \pm
8.9$~km~s$^{-1}$ differs from the single optically-measured value of
$8353 \pm 57$~km~s$^{-1}$ at the 6.8~$\sigma$ level. While the velocity
of the peak \HI emission is somewhat higher than the measured central
velocity, the discrepancy is still significant.

{\bf IRAS 21054+2314:}
This galaxy, with a redshift of $z=0.0487$, shows both \HI 21~cm
emission and absorption, and OH 1667~MHz absorption (Figures~1 and 3).
These detections have not been previously reported in the literature.
Its \HI absorption feature is at a higher velocity than the emission,
suggesting infalling gas towards the central continuum source.

{\bf IRAS 21442+0007:}
This source shows a previously unreported \HI 21~cm absorption line
at $cz = 22241$~km~s$^{-1}$ or $z=0.0742$. However, its \HI spectrum (Figure~1)
also shows a wide, previously unknown, emission line centered at
$cz=22555$~km~s$^{-1}$ or $z=0.0752$ with $\Delta V_{50\%} =
207.3$~km~s$^{-1}$. This emission does not seem to be arising from the
target source IRAS~21442+0007, but from the nearby galaxy
SDSS~J214651.97+002302.2 at $cz=22575 \pm 49$~km~s$^{-1}$ (or $z=0.0753
\pm 0.0002$).  The projected angular separation between the two sources
is 1.8~arcmin, meaning that SDSS~J214651.97+002302.2 was situated close
to the half-power point of the $\sim3.5$~arcmin L-band beam in our
observations. After correcting for the beam, we derive an \HI mass value
of $9 \times 10^9~M_ {\odot}$ for SDSS~J214651.97+002302.2.  The
proximity of IRAS 21442+0007 and SDSS~J214651.97+002302.2 in both
angular distance and velocity suggests that they may be members
of the same galaxy cluster or group.

{\bf IRAS 22045+0959:}
Also known as NGC 7212, this triple system has a cataloged redshift of
$z=0.0266$.  Optical studies show it to be a system of three
interacting galaxies lying within an area of radius
$\approx$0.5~arcmin.  NGC~7212~NED~1 is a small spiral of undetermined
redshift. NED~7212~NED~2 has an optical spectrum characteristic of a
Seyfert~2 nucleus, though showing a Seyfert~1-like spectrum in
polarized light  (e.g., \citealp{VVA93,WAS81}). Its optical radial
velocity is given as $7927 \pm  18$~km~s$^{-1}$ by \citet{FAL99},
but as 7800~km~s$^{-1}$ (with no formal error) by \citet{STR92}.
NED~7212~NED~3 appears to be interacting strongly with
NED~2, and has an optical radial velocity of $8167 \pm  74$~km~s$^{-1}$
\citep{FAL99}.

The IR luminosity of the system is $L_{\rm IR}=10^{11.15 \pm 0.10}
L_\odot$.  Our calculated \HI mass for the system is $10^{10.01 \pm
0.02} M_\odot$, slightly higher than the median ($10^{9.78} M_\odot$)
and the mean ($10^{9.69 \pm 0.08} M_\odot$) we find for sources with IR
luminosities in the range $10^{11}-10^{11.49} L_\odot$.  Our \HI
spectrum (Figure~2) may represent two spiral-galaxy-like, double-horned
spectra. The radial velocity of NED~3 agrees well with the higher
redshift feature.  However, while the \citet{STR92} velocity for
NED~2 would be compatible with the lower redshift feature, the \citet{FAL99}
velocity would place the systemic velocity near the ``central
gap'' between the two features. The depth of this gap between the two
features drops to slightly more than 3$\sigma$ below the continuum
level. Hence it could represent \HI absorption, while
also being compatible with a lack of \HI emission at that velocity.
Clearly, high resolution \HI synthesis mapping is needed to resolve the
actual situation in this system. Here, we have taken the \HI
classification of IRAS~22045+0959 to be a mixture of emission and
absorption (see Table~3), though we caution that this may need revision
following future observations.

{\bf IRAS 22523+3156:}
This object also displays previously unreported \HI and OH lines (Figures~1
and 3). IRAS 22523+3156 has a relatively low IR luminosity, and is at a
redshift of $z=0.0212$. Its \HI spectral shape is peculiar, with a deep
absorption feature in the middle of the emission line. Its OH spectrum
displays two nicely defined absorption features of the redshifted
1667 and 1665~MHz main lines.  The \HI and OH absorption lines arise
at the same velocity (see Tables~3 \& 5), indicating that they
originate from the same region.

The ratio of $\int{\tau dv}$ for the 1667 and 1665~MHz absorption
lines is $1.88 \pm 0.03$, comparable with the expected ratio in thermal
equilibrium of 1.8. In thermal equilibrium, the satellite lines should
be found with a hyperfine ratio relative to the 1667~MHz line of 0.111.
The measured rms level for the 1720 MHz satellite line (Table~5) gives
a 3$\sigma$ upper limit for this ratio of 0.131, indicating that our
non-detection is not inconsistent with thermal equilibrium. This is
especially the case as very low level RFI appears to be present near
this satellite line in some of the data. No rms value was estimated
for the 1612~MHz satellite line, because the spectrum was severely affected
by RFI.

{\bf IRAS 22595+1541:}
Our \HI spectrum for this target provides an estimated heliocentric
velocity of $2089.6 \pm 0.5$~km~s$^{-1}$, differing at the 7$\sigma$ level
from the weighted optically-measured value for the galaxy NGC~7465
(type~SB0) of $1977 \pm 16$~km~s$^{-1}$.  This galaxy is a member of
the galaxy group RSCG~83 for which NED provides a \HI velocity of
$2091 \pm 5$~km~s$^{-1}$ from the HIPASS \HI survey. Within our
3.5~arcmin HPBW we could potentially be seeing emission from three
galaxies, (a) NGC~7465; the prime target, (b) NGC~7464, an E1 pec
galaxy with an optical radial velocity of $1787 \pm 17$~km~s$^{-1}$,
and (c) NGC~7463, an SABb pec galaxy with an optical radial velocity of
$2439 \pm 24$~km~s$^{-1}$. The very broad \HI spectrum seen in Figure~2
suggests that in all probability we are seeing emission from all three,
with that of NGC~7465 dominating.

{\bf IRAS 23050+0359:}
This galaxy, at $z=0.0474$, is a new \HI emission detection, 
but has a previously known OH megamaser which we also detect
(Figures 1 and 4). It has a total IR luminosity of $L_{\rm IR}=10^{11.67
\pm 0.16} L_\odot$, making it one of the more IR luminous objects in
our sample.  The calculated \HI mass from its distorted spectrum is
$10^{9.71\pm 0.08} M_\odot$; a value that is slightly lower than the
median ($10^{9.86} M_\odot$) and the mean ($10^{9.99 \pm 0.10}
M_\odot$) for objects with IR luminosities greater than $10^{11.50}
L_\odot$ (Table~7).

{\bf IRAS 23204+0601:}
Also known as III Zw 103, this galaxy is at a redshift of $z=0.0560$,
and is also a new \HI emission detection (Figure 1).  With a
total IR luminosity of $L_{\rm IR}=10^{11.88 \pm 0.16} L_\odot$, it is also
one of the more IR luminous sources in our sample. We were unable to
determine if there is OH emission/absorption in this source, because
the spectra of all its main and satellite lines were severely
affected by RFI. Its \HI mass ($10^{9.76\pm 0.07} M_\odot$) is lower
than the median and mean of galaxies with comparable IR luminosities (Table~7).

{\bf IRAS 23327+2913:}
Both the \HI and OHM emission lines (Figures~1 and 3) of this ULIRG
($z=0.1067$) are new detections.  A study by \citet{DVT01} showed
that this object is a system of two interacting galaxies separated by
20~kpc. Its northern component is disturbed, while the southern component
is a normal spiral with a very thick bar structure.  High resolution
radio interferometric observations are needed to establish from which
galaxy in this system the \HI and OHM emission lines originate.

{\bf IRAS 23410+0228:}
This galaxy is at a redshift of $z=0.0912$, which is the second highest
redshift in our observed sample. It is one of two sources with no \HI
detection, with the other source being IRAS~20210+1121 ($z=0.0564$;
see above).  Its IR luminosity is $10^{12.08\pm 0.17}~L_{\odot}$ and the
derived 3$\sigma$ upper limit for its \HI mass, assuming a 400~km~s$^{-1}$
velocity width, is $10^{10.30}~M_{\odot}$. This \HI mass limit is
greater than the median and the mean \HI mass values for sources with
$L_{\rm IR} > 10^{11.50}~L_{\odot}$, which are $10^{9.86}~M_{\odot}$
and $10^{9.99 \pm 0.10}~M_{\odot}$, respectively. Therefore, the
non-detection of \HI emission from this source does not necessarily
imply hydrogen deficiency.

{\bf IRAS 23532+2513:}
This target consists of three galaxies lying within our HPBW at similar
redshifts. These are, (a) a starburst galaxy (type spiral:HII) lying
14~arcsec from our pointing position with an optical radial velocity of
$17584 \pm 69$~km~s$^{-1}$, (b) a peculiar Sy1 galaxy also 14 arcsec from
our pointing with an optical radial velocity of $17121 \pm
18$~km~s$^{-1}$, and (c) a spiral:HII galaxy about 1~arcmin from our
pointing with an optical radial velocity of $17273 \pm 68$~km~s$^{-1}$.
Our broad \HI spectrum (Figure~2) shows emission over the full range of
velocities covered by the 3 component galaxies.

\section{Acknowledgements}
We thank the anonymous referee for valuable comments and suggestions.
This research has made use of the NASA/IPAC Extragalactic Database
(NED) which is operated by the Jet Propulsion Laboratory, California
Institute of Technology, under contract with the National Aeronautics
and Space Administration.  We also acknowledge the usage of the
HyperLeda database (http://leda.univ-lyon1.fr). M.X.F. is grateful for
support from NAIC-Arecibo Observatory during a summer research
assistantship.

\begin{figure}
\epsscale{0.95}
\figurenum{1}
\plotone{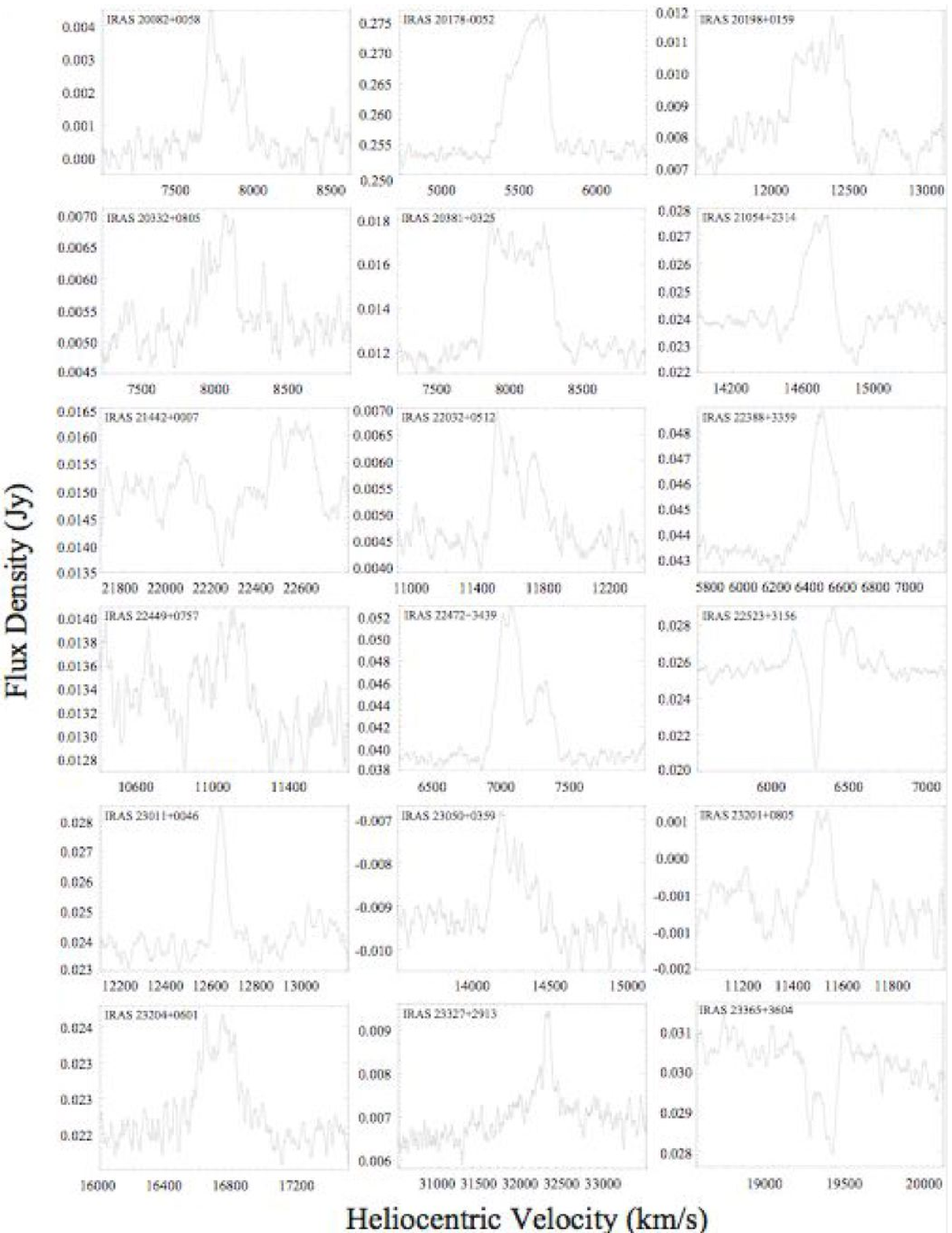}
\caption{Spectra of the new \HI 21~cm detections in our observed sample.}
\end{figure}

\begin{figure}
\epsscale{0.95}
\figurenum{2}
\plotone{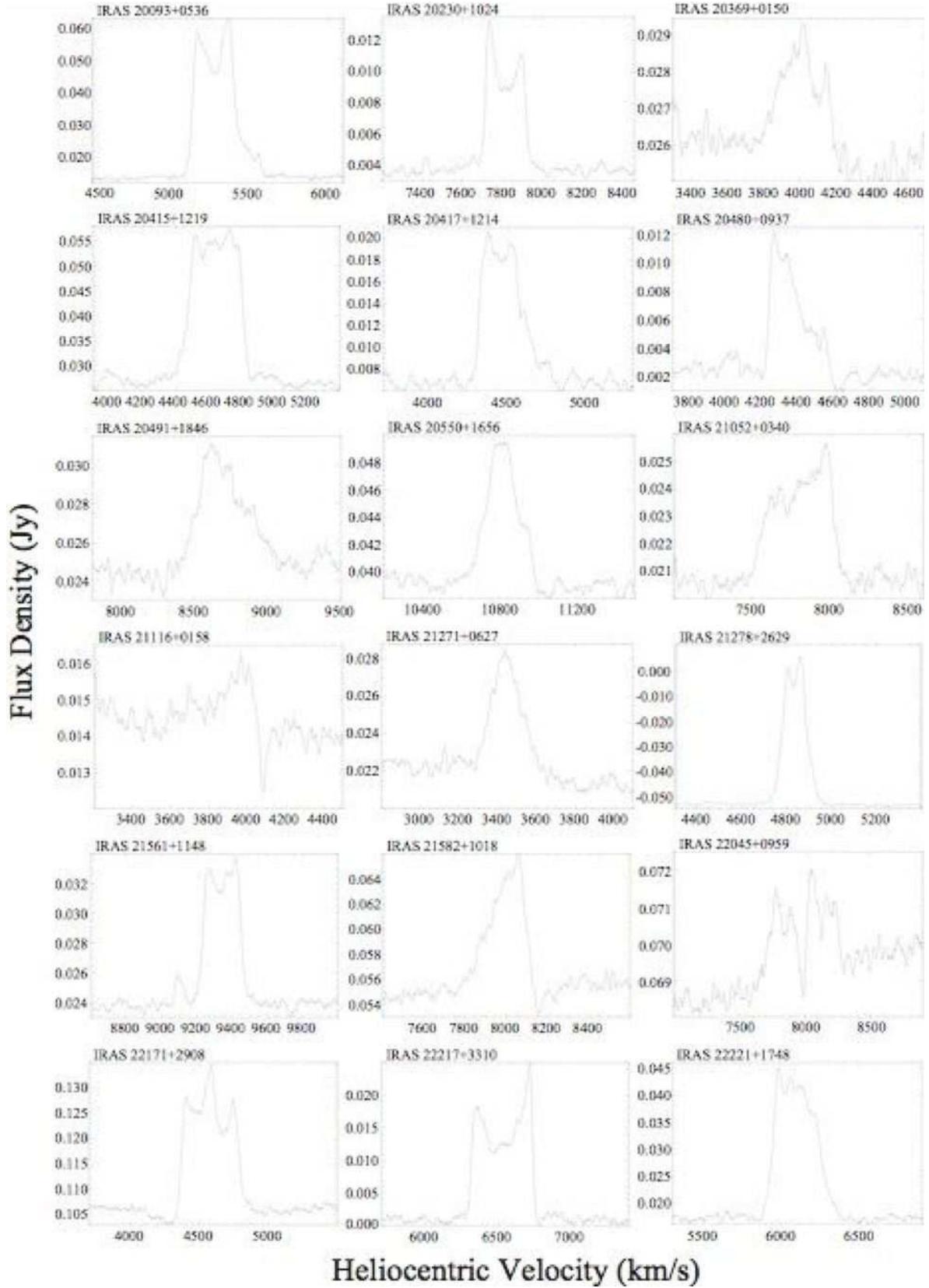}
\caption{Spectra of the previously known \HI 21~cm detections in our observed sample.}
\end{figure}

\begin{figure}
\epsscale{0.95}
\figurenum{2 cont}
\plotone{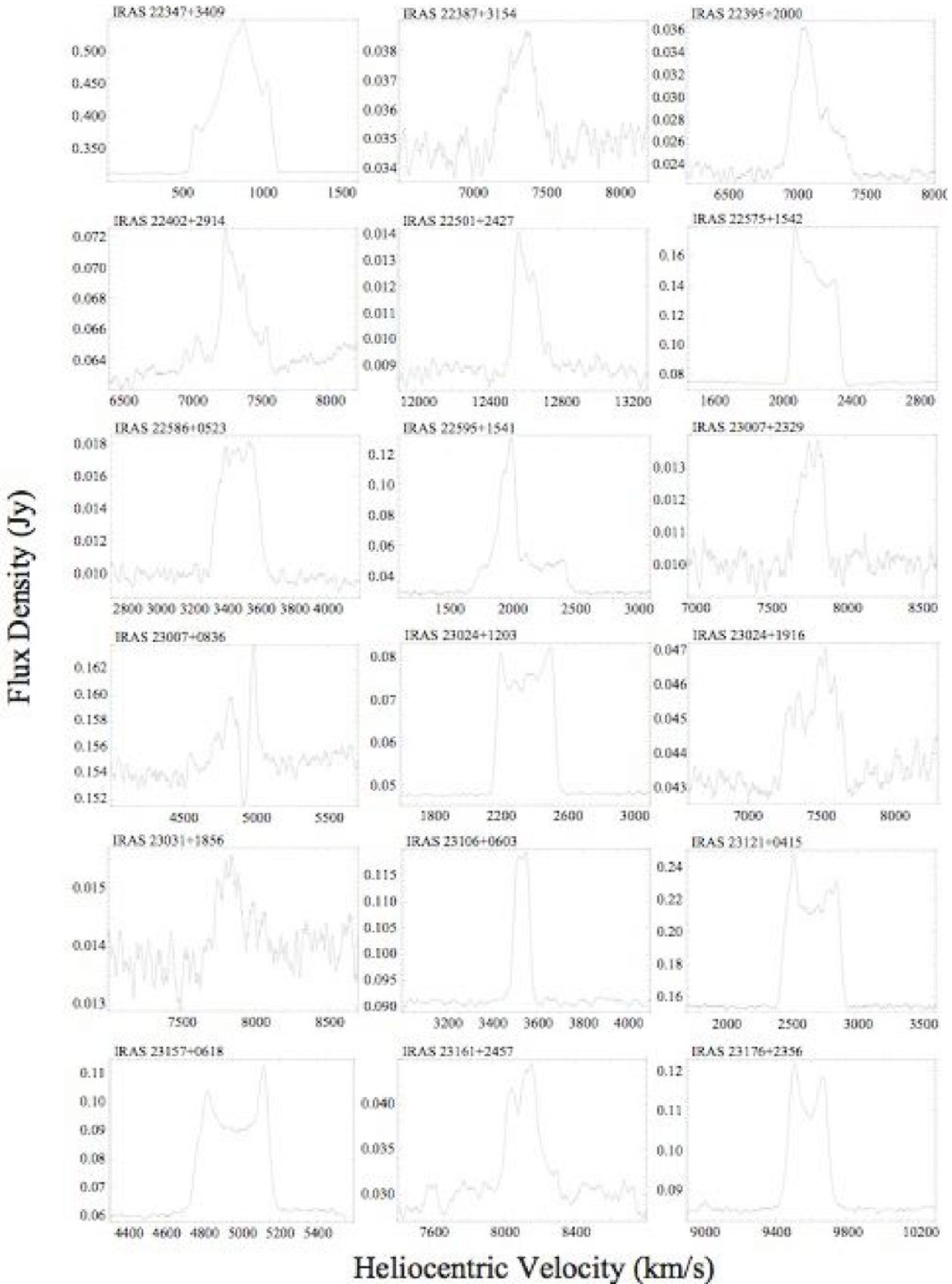}
\caption{}
\end{figure}

\begin{figure}
\epsscale{0.95}
\figurenum{2 cont}
\plotone{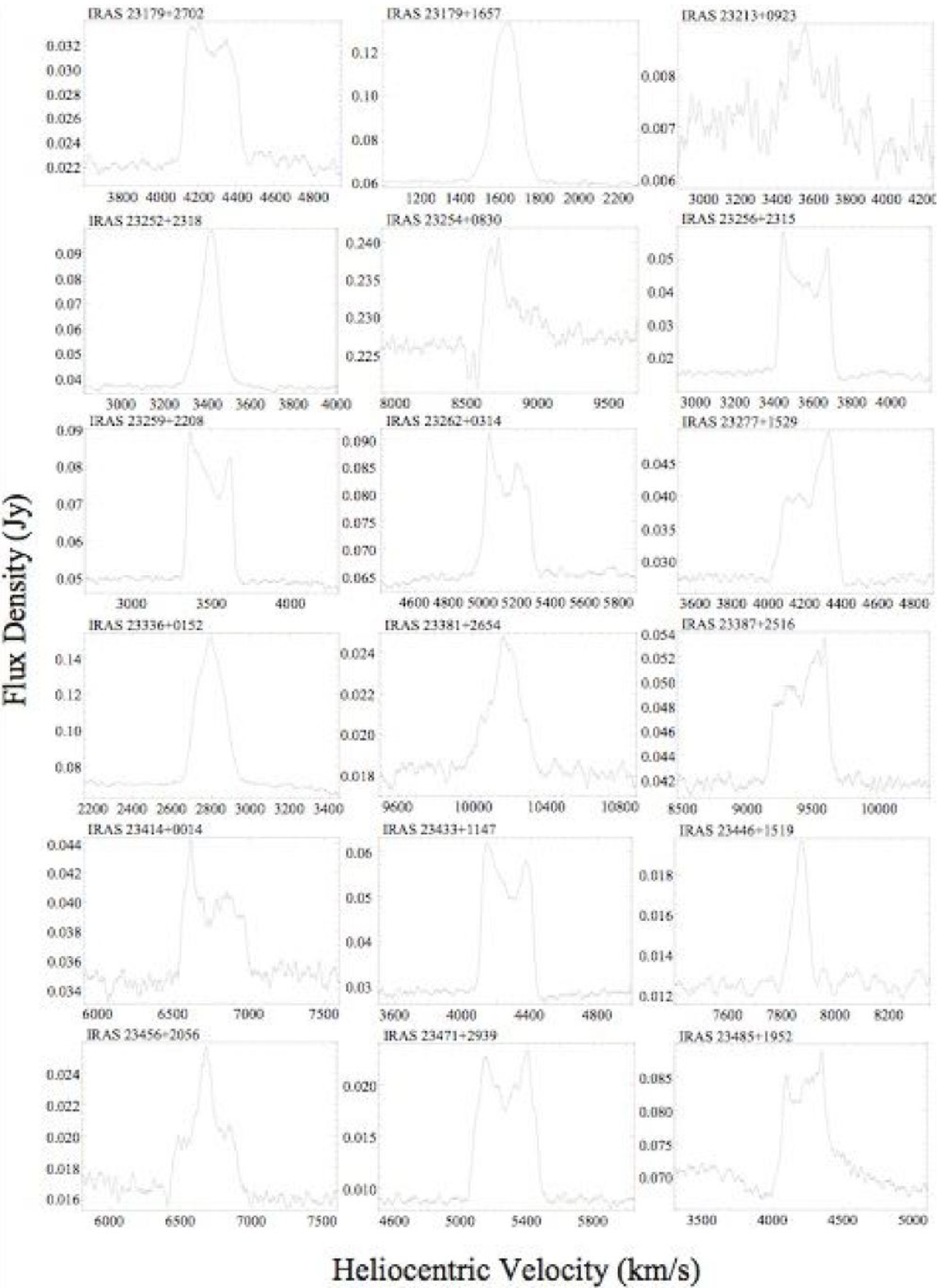}
\caption{}
\end{figure}

\begin{figure}
\epsscale{0.95}
\figurenum{2 cont}
\plotone{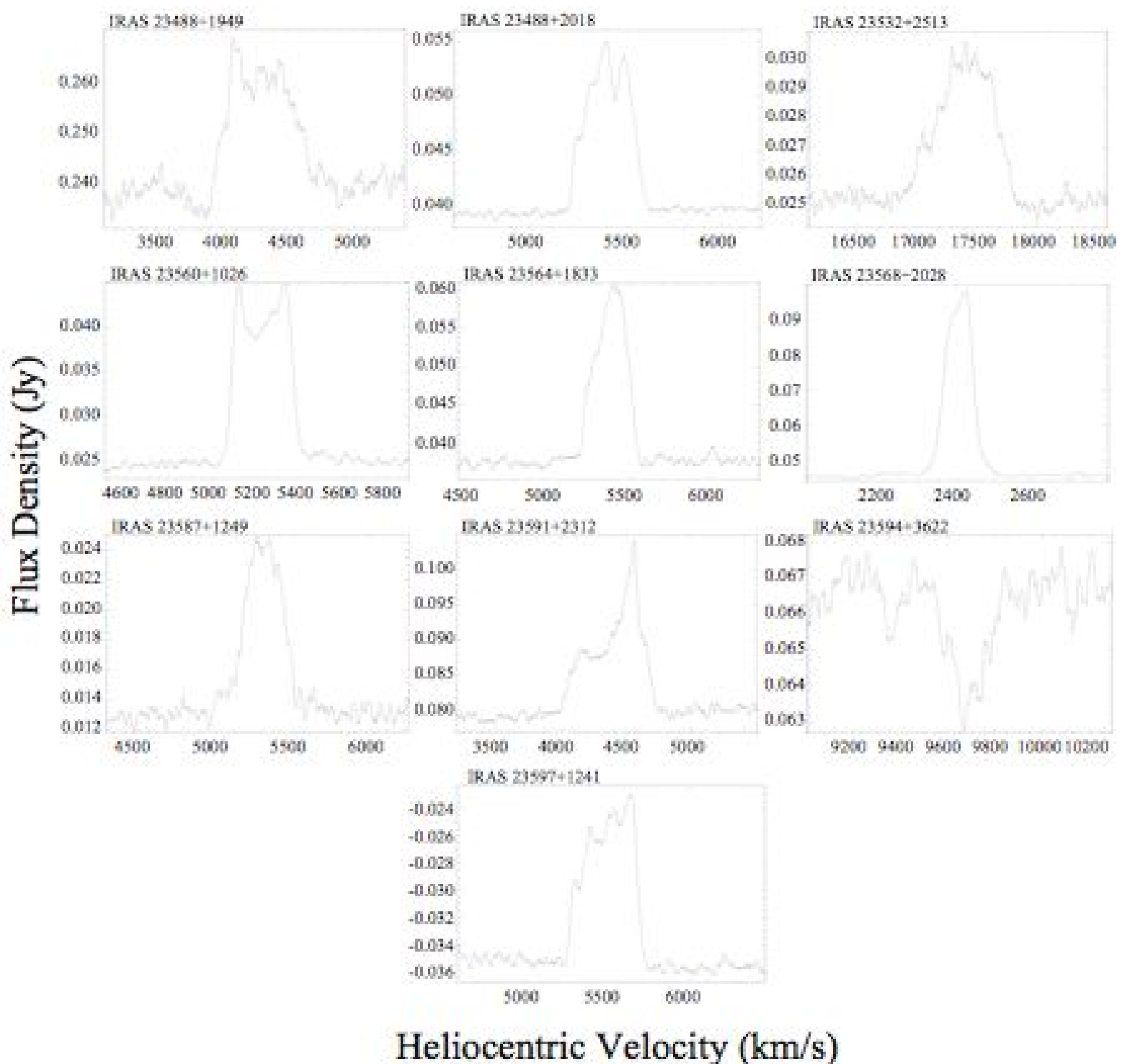}
\caption{}
\end{figure}

\begin{figure}
\figurenum{3}
\plotone{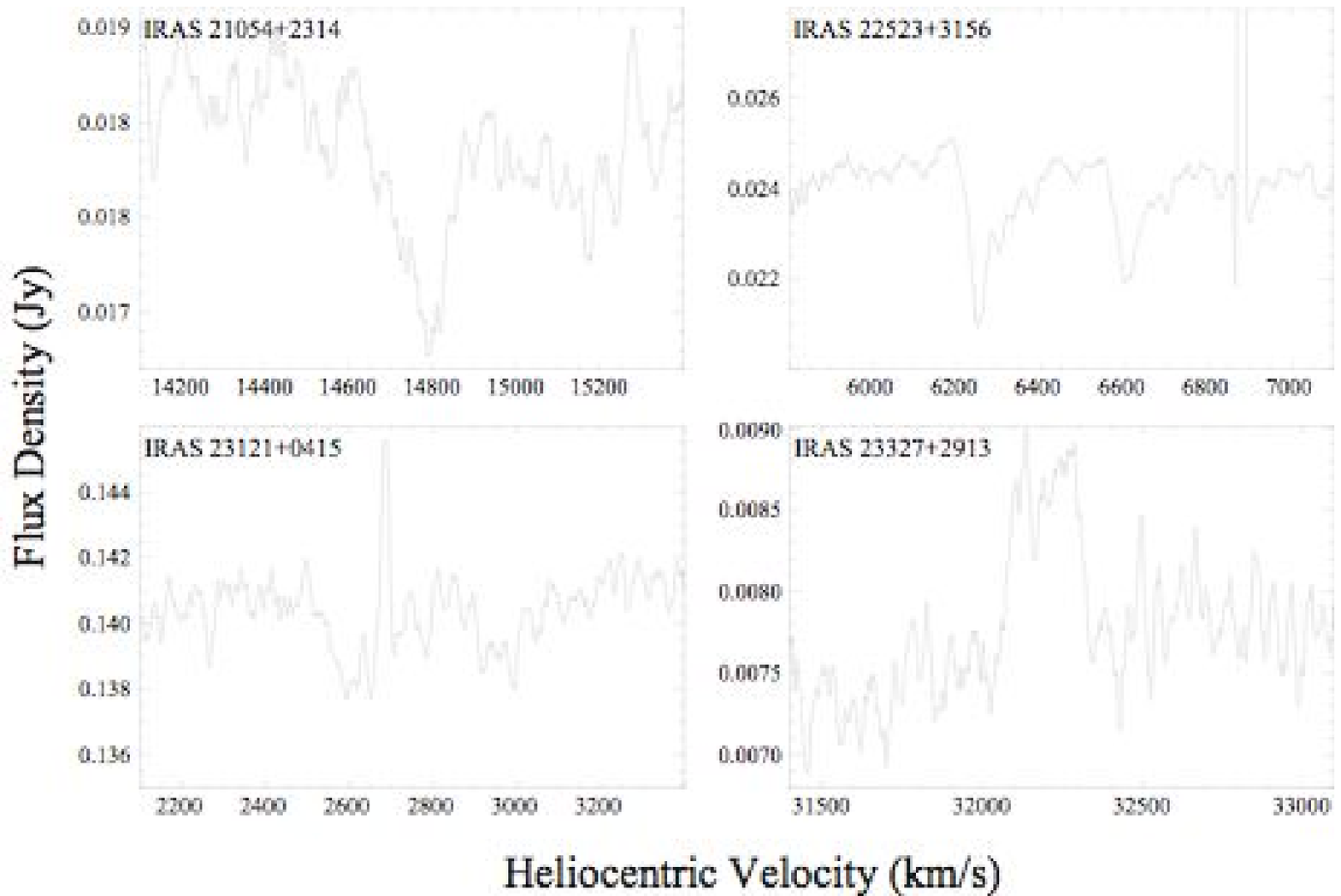}
\caption{Spectra of the new OH 18~cm main line detections; three are
in absorption, one in emission.  The spectral features at
6880~km~s$^{-1}$ in the spectrum of IRAS~22523+3156 and at
2685~km~s$^{-1}$ in the spectrum of IRAS~23121+0415 are due to RFI.
The spectra used the rest frequency of the 1667.359~MHz line 
for the velocity scale.}
\end{figure}

\begin{figure}
\figurenum{4}
\epsscale{0.9}
\plotone{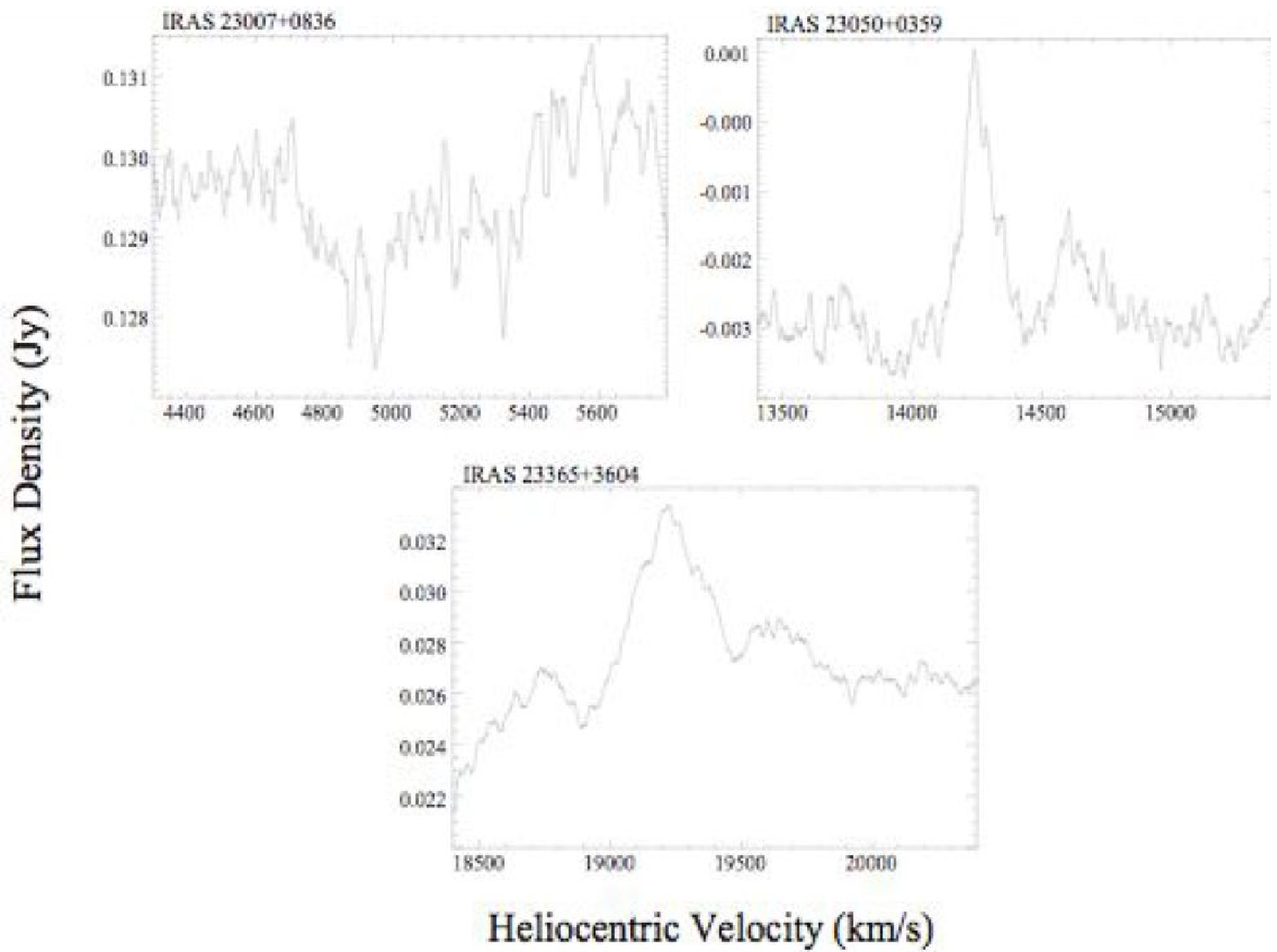}
\caption{Spectra of the previously known OH 18~cm main line detections in our observed sample.
The spectra used the rest frequency of the 1667.359~MHz line for the velocity scale.}
\end{figure}

\begin{figure}
\figurenum{5}
\epsscale{0.8}
\plotone{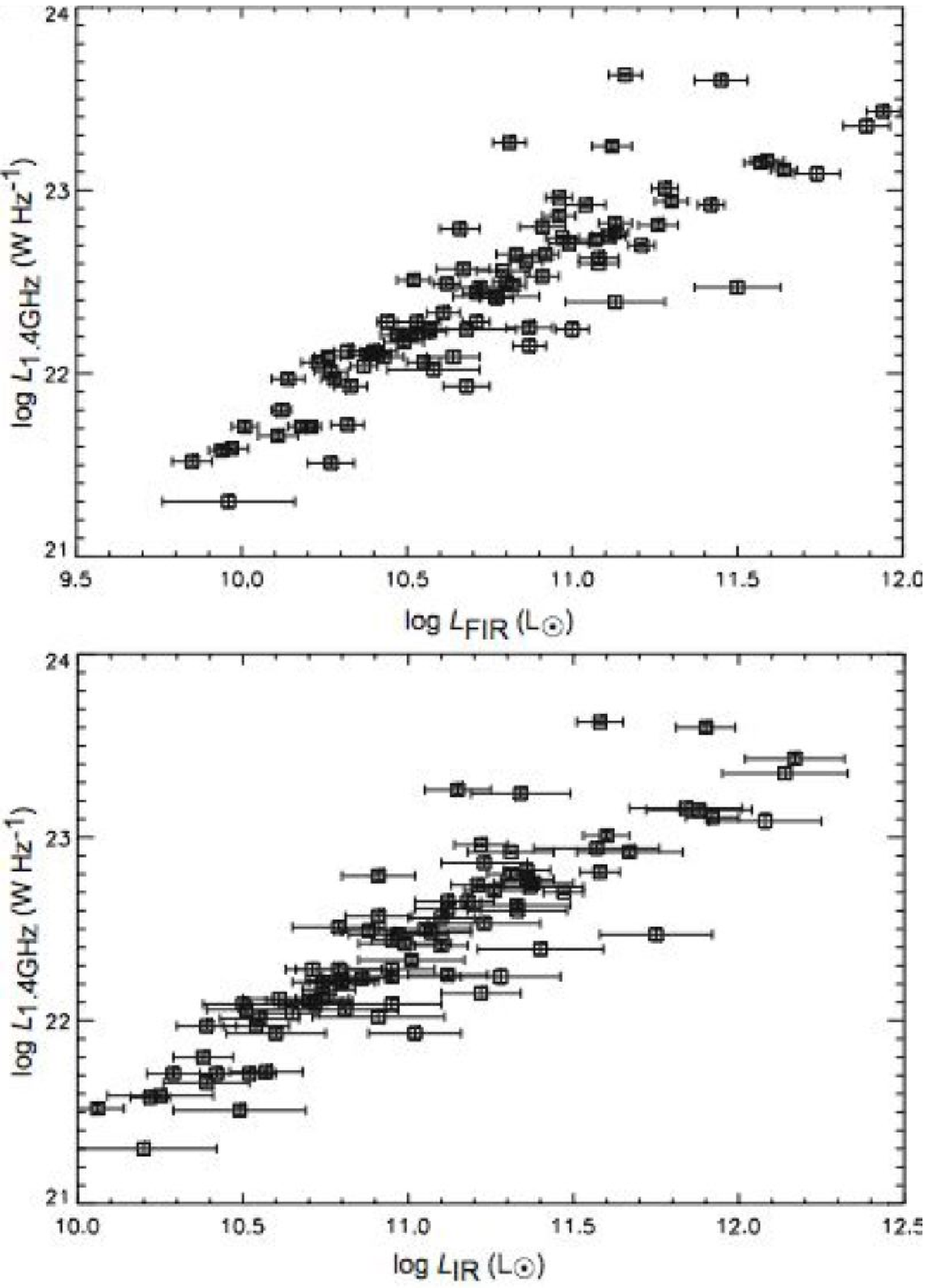}
\caption{Logarithmic plots of the 1.4~GHz radio luminosity vs.~FIR luminosity ({\it top})
and total IR luminosity ({\it bottom}) of the galaxies in our observed sample.
The correlation coefficients are 88\% and 89\% for the {\it top} and {\it bottom} plots,
respectively.}
\end{figure}

\begin{figure}
\figurenum{6}
\epsscale{0.8}
\plotone{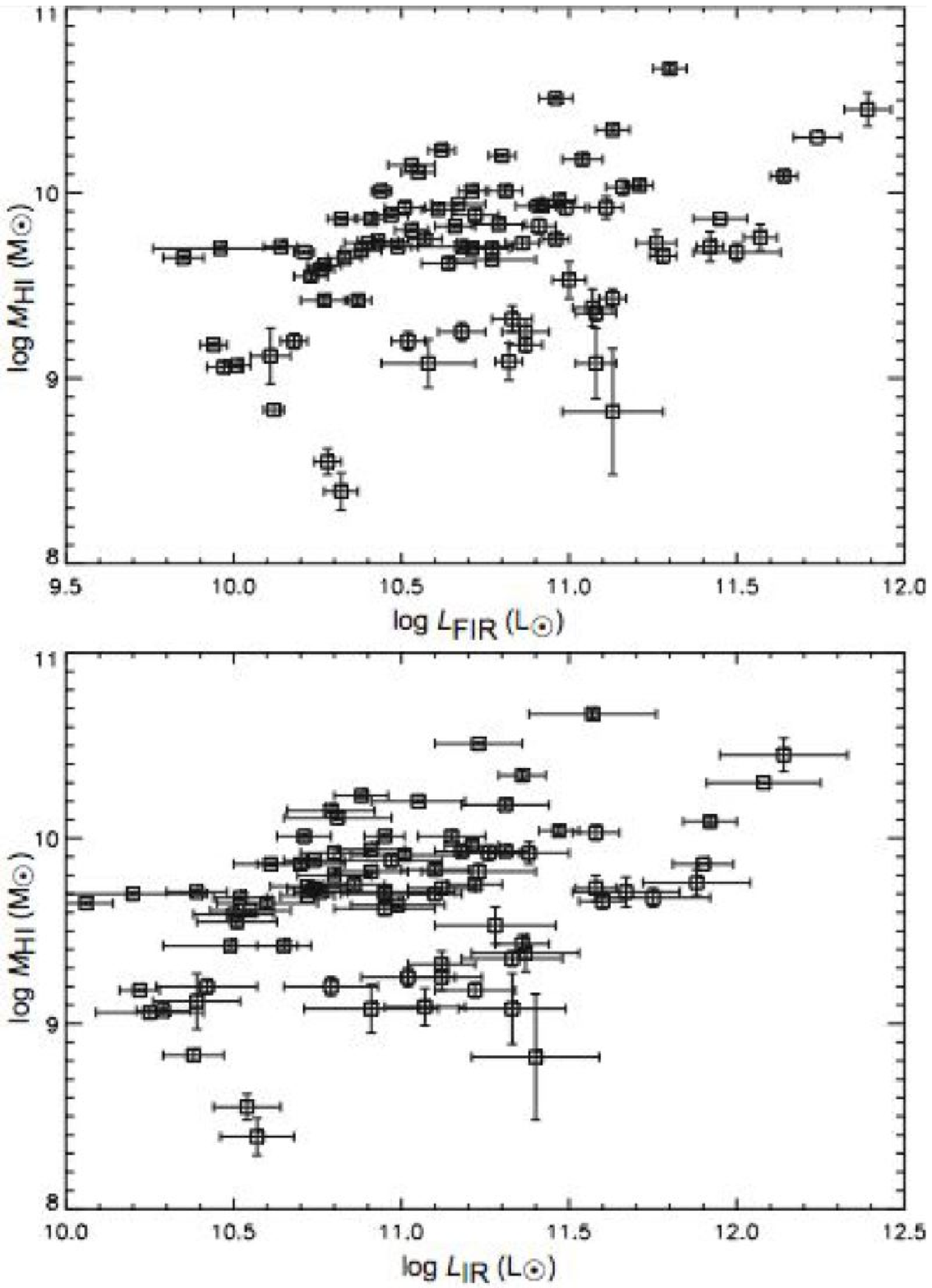}
\caption{Logarithmic plots of the \HI mass vs.~FIR luminosity ({\it top}) and total IR luminosity ({\it bottom})
of the galaxies in our observed sample.
The correlation coefficient in each of these plots is 42\%.}
\end{figure}

\begin{figure}
\figurenum{7}
\epsscale{0.8}
\plotone{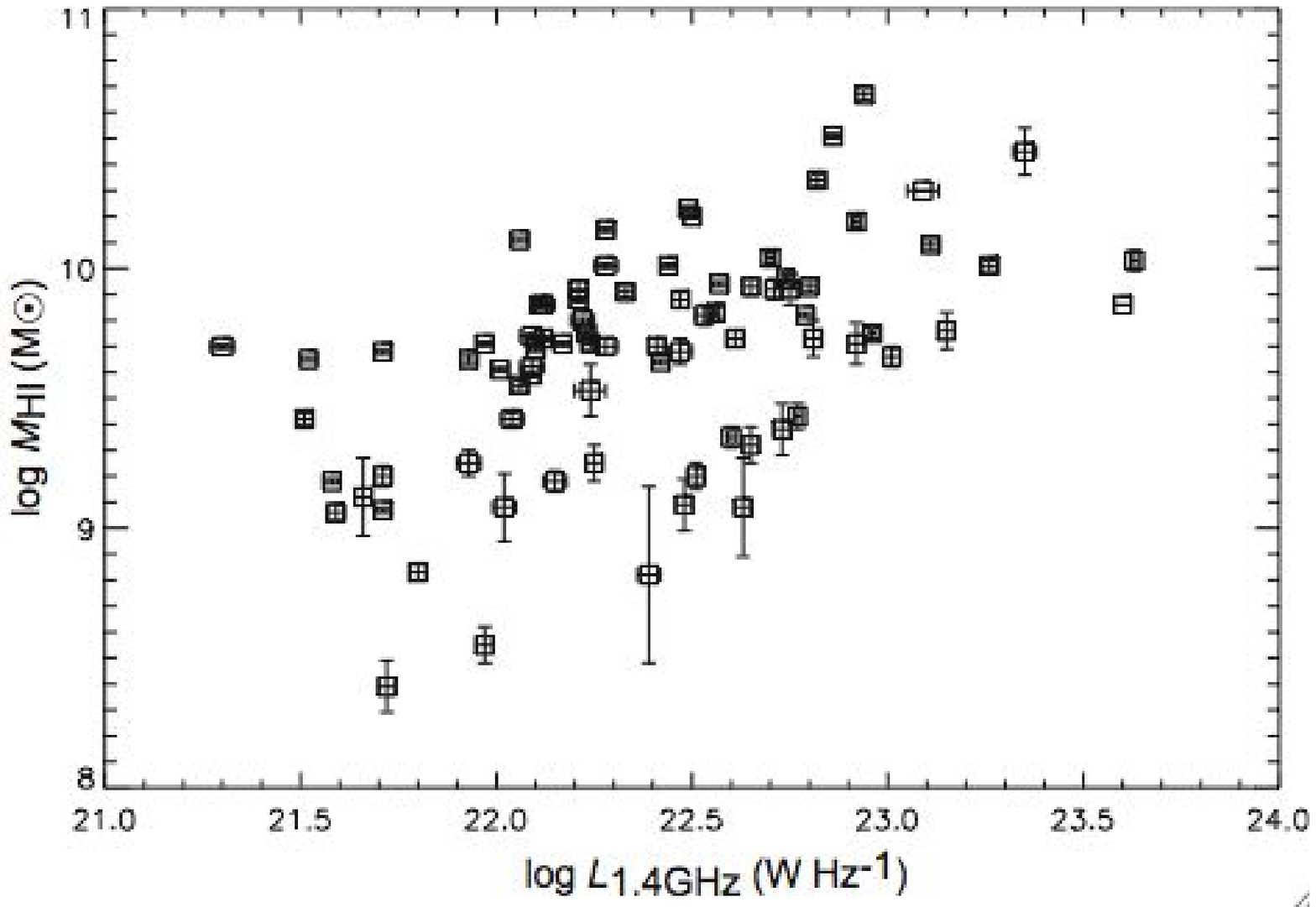}
\caption{Logarithmic plot of the \HI mass vs.~the 1.4~GHz radio luminosity of the galaxies in our observed sample.
The correlation coefficient in this plot is 53\%.}
\end{figure}

\clearpage

\begin{deluxetable}{cccccccc}
\tabletypesize{\footnotesize}
\rotate
\tablenum{1}
\tablecolumns{8}
\tablewidth{0pc}
\tablecaption{Sample Properties}
\tablehead{
\colhead{IRAS Name}
& \colhead{Other Names}
& \colhead{R.~A.~(J2000)}
& \colhead{Decl.~(J2000)}
& \colhead{$z$}
& \colhead{$S_{60{\mu}m}$ }
& \colhead{$S_{\rm 1.4GHz}$}
& \colhead{Morphology} \\
\colhead{}
& \colhead{}
& \colhead{}
& \colhead{}
& \colhead{}
& \colhead{(Jy)}
& \colhead{(mJy)}
& \colhead{} \\
\colhead{(1)}
& \colhead{(2)}
& \colhead{(3)}
& \colhead{(4)}
& \colhead{(5)}
& \colhead{(6)}
& \colhead{(7)}
& \colhead{(8)}}
\startdata
20082$+$0058 &                                            & 20 10 46.29   & +01 07 13.6   &  0.0258  &   2.69 &   5.6 &  S?                          \\
20093$+$0536 & UGC 11522                                  & 20 11 49.36   & +05 45 47.0   &  0.0175  &   3.96 &  44.9 &  Sbc                         \\
20178$-$0052 &                                            & 20 20 28.00   &$-00$ 42 36.5  &  0.0185  &   3.08 &  21.9 &  HII                         \\
20198$+$0159 &                                            & 20 22 23.07   & +02 09 19.1   &  0.0414  &   2.40 &  14.6 &  S?                          \\
20210$+$1121 &                                            & 20 23 25.54   & +11 31 37.2   &  0.0564  &   3.38 &  53.7 &  Sy2/Sc                    \\
20230$+$1024 &                                            & 20 25 30.64   & +10 34 21.5   &  0.0260  &   2.01 &   8.1 &  Sbc                         \\
20332$+$0805 &                                            & 20 35 39.16   & +08 16 15.8   &  0.0279  &   3.11 &  19.2 &  HII/S?                      \\
20369$+$0150 &                                            & 20 39 26.45   & +02 01 04.1   &  0.0129  &   2.41 &  11.5 &  Sb/S0-a                   \\
20381$+$0325 &                                            & 20 40 39.34   & +03 35 47.7   &  0.0259  &   2.07 &  18.5 &                            \\
20415$+$1219 & NGC 6956, UGC 11619                        & 20 43 53.46   & +12 30 35.4   &  0.0156  &   3.18 &  35.9 &  SBb                         \\
20417$+$1214 & UGC 11620                                  & 20 44 09.74   & +12 25 05.0   &  0.0149  &   2.55 &  23.4 &  Sb/S0-a                 \\
20480$+$0937 &                                            & 20 50 29.09   & +09 49 05.5   &  0.0147  &   2.34 &  10.9 &  SABa                        \\
20491$+$1846 & UGC 11643                                  & 20 51 25.90   & +18 58 04.8   &  0.0291  &   2.79 &  23.4 &  SBb                         \\
20550$+$1656 & II Zw 096                                  & 20 57 24.14   & +17 07 41.2   &  0.0361  &  13.30 &  43.2 &  HII/S0-a                  \\
21052$+$0340 & UGC 11680                                  & 21 07 45.88   & +03 52 40.5   &  0.0259  &   2.80 &  16.9 &  Scd/HII Sy2/E-S0        \\
21054$+$2314 &                                            & 21 07 43.36   & +23 27 06.4   &  0.0487  &   2.23*&  11.6 &  S?                          \\
21116$+$0158 & IC 1368, UGC 11703                         & 21 14 12.57   & +02 10 41.2   &  0.0130  &   4.03 &  25.8 &  Sa;Sy2                     \\
21271$+$0627 & NGC 7074, II Zw 133                        & 21 29 38.93   & +06 40 57.3   &  0.0116  &   3.11 &  21.7 &  S?                          \\
21278$+$2629 & NGC 7080, UGC 11756                        & 21 30 01.81   & +26 43 05.3   &  0.0161  &   3.16 &  28.4 &  SBb                         \\
21396$+$3623 &                                            & 21 41 41.65   & +36 36 47.4   &  0.1493  &   2.16*&  13.8 &                              \\
21442$+$0007 &                                            & 21 46 51.28   & +00 21 13.5   &  0.0740  &   2.11 &  10.9 &  HII                       \\
21561$+$1148 & Mrk 0518, UGC 11865                        & 21 58 36.09   & +12 02 19.4   &  0.0311  &   2.73*&  29.1 &  I?/S?                       \\
21582$+$1018 & Mrk 0520, UGC 11871                        & 22 00 41.40   & +10 33 07.5   &  0.0266  &   4.15 &  57.9 &  Sb;Sy1.9                  \\
22032$+$0512 &                                            & 22 05 47.08   & +05 27 16.3   &  0.0383  &   2.43 &   5.1 &  S?                          \\
22045$+$0959 & NGC 7212                                   & 22 07 02.05   & +10 14 02.7   &  0.0266  &   2.88 & 114.1 &  Sab;Sy2                 \\
22171$+$2908 & NGC 7253A, Arp 278, UGC 11984              & 22 19 27.94   & +29 23 41.6   &  0.0152  &   5.86 &  72.1 &  SABc                      \\
22217$+$3310 & UGC 12022                                  & 22 24 02.76   & +33 26 08.9   &  0.0218  &   2.25 &  10.8 &  Sbc                     \\
22221$+$1748 &                                            & 22 24 33.40   & +18 03 56.5   &  0.0204  &   2.66 &  21.0 &  S-I/Sc                    \\
22347$+$3409 & NGC 7331, UGC 12113                        & 22 37 04.67   & +34 24 28.0   &  0.0027  &  23.10 & 217.6 &  SAb;LIN/Sbc               \\
22387$+$3154 & Mrk 0917, UGC 12149                        & 22 41 07.60   & +32 10 11.1   &  0.0244  &   3.71 &  33.3 &  SBa;Sy2                   \\
22388$+$3359 & UGC 12150                                  & 22 41 12.28   & +34 14 57.4   &  0.0214  &   8.17 &  38.3 &  SB0-a;HII/LIN             \\
22395$+$2000 &                                            & 22 41 56.03   & +20 15 41.6   &  0.0233  &   2.56 &  16.9 &  Sy2;HII/S0                  \\
22402$+$2914 &                                            & 22 42 38.53   & +29 30 23.2   &  0.0244  &   2.02 &  13.4 &  S?                        \\
22449$+$0757 &                                            & 22 47 28.17   & +08 13 37.6   &  0.0372  &   2.87 &  14.0 &  S?                          \\
22472$+$3439 &                                            & 22 49 32.17   & +34 55 09.2   &  0.0234  &   4.98 &  43.9 &  Sbrst;LIN                   \\
22501$+$2427 & Mrk 0309, IV Zw 121                        & 22 52 34.76   & +24 43 44.8   &  0.0421  &   3.43 &   7.3 &  Sa;Sbrst/Sy2                \\
22523$+$3156 &                                            & 22 54 45.05   & +32 12 47.8   &  0.0212  &   2.27 &  32.0 &  Sb/Sbc                      \\
22575$+$1542 & NGC 7448,  Arp 013, UGC 12294              & 23 00 03.60   & +15 58 50.8   &  0.0073  &   7.23 &  81.5 &  SAbc/Sc                   \\
22586$+$0523 & UGC 12304                                  & 23 01 08.21   & +05 39 15.8   &  0.0116  &   2.06 &  13.5 &  Sc                        \\
22595$+$1541 & NGC 7465, Mrk 0313, UGC 12317              & 23 02 00.93   & +15 57 51.2   &  0.0066  &   3.80 &  19.1 &  SB0;Sy2                   \\
23007$+$2329 &                                            & 23 03 09.28   & +23 45 32.3   &  0.0259  &   3.64 &  11.9 &  S?                        \\
23007$+$0836 & NGC 7469, Arp 298, Mrk 1514, UGC 12332     & 23 03 15.62   & +08 52 26.1   &  0.0163  &  25.90 & 180.5 &  SABa/Sy1.2                \\
23011$+$0046 &                                            & 23 03 41.29   & +01 02 38.0   &  0.0420  &   2.58 &  13.3 &  Sm                        \\
23024$+$1203 & NGC 7479, UGC 12343                        & 23 04 56.63   & +12 19 20.6   &  0.0079  &  12.80 &  99.0 &  SBc;LIN/SBbc;Sy1.9      \\
23024$+$1916 &                                            & 23 04 56.61   & +19 33 08.1   &  0.0251  &   7.53 &  42.6 &  LIN/Sa                  \\
23031$+$1856 &                                            & 23 05 36.16   & +19 12 29.6   &  0.0262  &   2.09 &   6.8 &  Sbrst/Sa                  \\
23050$+$0359 &                                            & 23 07 35.73   & +04 15 59.8   &  0.0474  &   3.89 &  15.8 &  HII/S?                      \\
23106$+$0603 & NGC 7518, Mrk 0527, UGC 12422              & 23 13 12.67   & +06 19 23.3   &  0.0118  &   4.20 &  10.7 &  SABa                      \\
23121$+$0415 & NGC 7541, UGC 12447                        & 23 14 44.00   & +04 32 00.8   &  0.0090  &  19.30 & 162.4 &  SBbc;HII                  \\
23157$+$0618 & NGC 7591, UGC 12486                        & 23 18 16.32   & +06 35 08.5   &  0.0165  &   7.22 &  52.1 &  SBbc;Sy;LIN                 \\
23161$+$2457 & Mrk 0319, UGC 12490                        & 23 18 38.41   & +25 13 58.4   &  0.0270  &   4.27 &  31.6 &  SBa;Sbrst                 \\
23176$+$2356 & NGC 7620, Mrk 0321, UGC 12520              & 23 20 05.65   & +24 13 15.9   &  0.0320  &   2.41 &  31.5 &  Scd;Sbrst/HII             \\
23179$+$2702 & NGC 7624, Mrk 0323, UGC 12527              & 23 20 22.69   & +27 18 55.7   &  0.0143  &   3.16 &  24.6 &  SBc;HII/Sbrst             \\
23179$+$1657 & NGC 7625,  Arp 212, III Zw 102, UGC 12529  & 23 20 30.08   & +17 13 32.4   &  0.0054  &   9.33 &  60.3 &  SAa/HII                    \\
23201$+$0805 &                                            & 23 22 43.92   & +08 21 34.7   &  0.0378  &   2.33 &   7.4 &  Sy2/S?                     \\
23204$+$0601 & III Zw 103                                 & 23 23 01.60   & +06 18 05.8   &  0.0560  &   4.23 &  19.4 &  HII/S?                      \\
23213$+$0923 & NGC 7648, IC 1486, Mrk 0531, UGC 12575     & 23 23 53.86   & +09 40 02.4   &  0.0119  &   4.84 &  16.9 &  S0                          \\
23252$+$2318 & NGC 7673, Mrk 0325, IV Zw 149, UGC 12607   & 23 27 41.28   & +23 35 22.5   &  0.0114  &   4.91 &  43.4 &  SAc;HII;Sbrst               \\
23254$+$0830 & NGC 7674, Arp 182, Mrk 0533, UGC 12608     & 23 27 56.70   & +08 46 43.2   &  0.0289  &   5.59 & 220.9 &  SAbc;HII;Sy2                \\
23256$+$2315 & NGC 7677, Mrk 0326, UGC 12610              & 23 28 06.22   & +23 31 52.1   &  0.0119  &   3.96 &  16.8 &  SABbc;Sbrst                 \\
23259$+$2208 & NGC 7678, Arp 028, UGC 12614               & 23 28 27.31   & +22 25 07.3   &  0.0116  &   6.59 &  49.5 &  SABc;Sbrst Sy2              \\
23262$+$0314 & NGC 7679, Arp 216, Mrk 0534, UGC 12618     & 23 28 46.73   & +03 30 41.3   &  0.0171  &   7.41*&  55.8 &  SB0;HII;Sy1/SB0-a;Sy2   \\
23277$+$1529 & UGC 12633                                  & 23 30 13.57   & +15 45 40.6   &  0.0141  &   3.11 &  23.7 &  SB/Sbab                   \\
23327$+$2913 &                                            & 23 35 11.88   & +29 30 00.3   &  0.1067  &   2.10 &   7.8 &  LIN                         \\
23336$+$0152 & NGC 7714, Arp 284, Mrk 0538, UGC 12699     & 23 36 14.12   & +02 09 18.1   &  0.0093  &  10.40 &  65.8 &  SBb;HII;LIN              \\
23365$+$3604 &                                            & 23 39 01.24   & +36 21 09.0   &  0.0645  &   7.09 &  27.2 &  SBa;LIN                     \\
23381$+$2654 &                                            & 23 40 42.78   & +27 10 40.9   &  0.0339  &   2.17 &  13.1 &  Sa/S0                   \\
23387$+$2516 &                                            & 23 41 16.13   & +25 33 03.7   &  0.0314  &   3.02 &  37.3 &  Sb                      \\
23410$+$0228 &                                            & 23 43 39.65   & +02 45 06.1   &  0.0912  &   2.28 &   6.0 &  Sy1/S?                      \\
23414$+$0014 & NGC 7738, UGC 12757                        & 23 44 02.01   & +00 30 59.5   &  0.0226  &   4.47 &  36.2 &  SBb                       \\
23433$+$1147 & IC 1508, UGC 12773                         & 23 45 55.04   & +12 03 42.6   &  0.0142  &   3.28 &  30.1 &  Sdm/Scd                     \\
23446$+$1519 &                                            & 23 47 09.40   & +15 35 49.4   &  0.0259  &   4.26 &   9.2 &  HII;Sy2/SBab                \\
23456$+$2056 & UGC 12787                                  & 23 48 13.77   & +21 13 03.5   &  0.0222  &   2.29 &  17.3 &  Sbc                       \\
23471$+$2939 & UGC 12798                                  & 23 49 39.73   & +29 55 55.1   &  0.0176  &   2.47 &  18.1 &  S?/Sc                       \\
23485$+$1952 & NGC 7769, Mrk 9005, UGC 12808              & 23 51 04.02   & +20 09 00.7   &  0.0140  &   4.34 &  59.9 &  SAb;HII;LIN                 \\
23488$+$1949 & NGC 7771, Mrk 9006, UGC 12815              & 23 51 24.90   & +20 06 41.3   &  0.0143  &  19.00 & 141.4 &  SBa;HII/Sbrst               \\
23488$+$2018 & Mrk 0331, UGC 12812                        & 23 51 26.79   & +20 35 10.6   &  0.0185  &  18.60 &  70.7 &  HII;Sy2/Sa                \\
23532$+$2513 &                                            & 23 55 49.99   & +25 30 21.9   &  0.0571  &   1.44 &  11.0 &  HII                       \\
23560$+$1026 & NGC 7794, UGC 12872                        & 23 58 34.09   & +10 43 42.2   &  0.0176  &   3.26 &  24.8 &  S?/Sbc                      \\
23564$+$1833 & UGC 12879                                  & 23 59 01.32   & +18 50 05.0   &  0.0180  &   2.64 &  18.2 &  S?/Sc                   \\
23568$+$2028 & NGC 7798, Mrk 0332, UGC 12884              & 23 59 25.60   & +20 45 00.1   &  0.0080  &   4.87 &  36.5 &  SBc/Sbrst                   \\
23587$+$1249 & NGC 7803, UGC 12906                        & 00 01 19.87   & +13 06 40.5   &  0.0179  &   2.02 &  12.3 &  S0-a                        \\
23591$+$2312 & III Zw 125                                 & 00 01 40.44   & +23 29 34.0   &  0.0145  &   6.13 & 131.9 &  SBc;Sbrst;LIN               \\
23594$+$3622 &                                            & 00 01 58.39   & +36 38 56.3   &  0.0321  &   4.48 &  75.0 &  Sy2/S0-a                    \\
23597$+$1241 & NGC 7810, UGC 12919                        & 00 02 19.08   & +12 58 18.0   &  0.0185  &   3.39 &  23.0 &  S0                          \\
\enddata
\tablenotetext{*}{The 60~$\mu$m values taken from the IRAS Point Source Catalogue (PSC) instead of the IRAS
Faint Source Catalogue (FSC).}
\end{deluxetable}

\begin{deluxetable}{lccccccccc}
\tabletypesize{\scriptsize}
\rotate
\tablenum{2}
\tablecolumns{10}
\tablewidth{0pc}
\tablecaption{Parameters of Galaxies with \HI Emission}
\tablehead{
\colhead{IRAS Name}
& \colhead{$\Delta t$}
& \colhead{rms}
& \colhead{$V_{\rm HI}$}
& \colhead{$\Delta{V_{50}}$}
& \colhead{$\int{Sdv}$}
& \colhead{$D_{\rm L}$}
& \colhead{log $\frac{L_{\rm FIR}}{\rm L_ {\odot}}$}
& \colhead{log $\frac{L_{\rm IR}}{\rm L_ {\odot}}$}
& \colhead{log $\frac{M_{\rm HI}}{\rm M_ {\odot}}$}\\
\colhead{}
& \colhead{(min)}
& \colhead{(mJy)}
& \colhead{(km $\rm s^{-1}$)}
& \colhead{(km $\rm s^{-1}$)}
& \colhead{(Jy km $\rm s^{-1}$)}
& \colhead{(Mpc)}
& \colhead{}
& \colhead{}
& \colhead{} \\
\colhead{(1)}
& \colhead{(2)}
& \colhead{(3)}
& \colhead{(4)}
& \colhead{(5)}
& \colhead{(6)}
& \colhead{(7)}
& \colhead{(8)}
& \colhead{(9)}
& \colhead{(10)}}
\startdata
20082$+$0058                 & 15   &  0.30   & $ 7784.6\pm 3.5 $&  $ 275.1\pm 7.0 $& $ 0.6\pm0.1 $ &  $ 111.9\pm0.1  $ &  $10.68\pm0.07$& $11.02\pm0.14$ & $9.25 \pm0.05 $  \\
20093$+$0536                 & 15   &  0.33   & $ 5286.5\pm 0.8 $&  $ 323.5\pm 1.6 $& $12.6\pm0.1 $ &  $  75.5\pm<0.1 $ &  $10.62\pm0.04$& $10.88\pm0.08$ & $10.23\pm<0.01$  \\
20178$-$0052$^{\diamond}$    &  5   &  0.90   & $ 5520.0\pm 3.2 $&  $ 317.6\pm 6.4 $& $ 5.6\pm0.4 $ &  $  78.9\pm0.1  $ &  $10.51\pm0.06$& $10.80\pm0.10$ & $9.92 \pm0.03 $  \\
20198$+$0159                 & 20   &  0.37   & $12301.0\pm 3.6 $&  $ 394.8\pm 7.3 $& $ 1.1\pm0.1 $ &  $ 178.8\pm0.1  $ &  $11.11\pm0.05$& $11.38\pm0.12$ & $9.92 \pm0.06 $  \\
20210$+$1121*                & 10   &  0.41   & $16905.0\pm 25.0$&  $ 400.0        $& $ 0.5       $ &  $ 248.7\pm0.4  $ &  $11.45\pm0.08$& $11.90\pm0.09$ & $< 9.86       $  \\
20230$+$1024                 & 15   &  0.27   & $ 7811.3\pm 2.4 $&  $ 222.1\pm 4.7 $& $ 1.4\pm0.1 $ &  $ 112.3\pm<0.1 $ &  $10.64\pm0.08$& $10.95\pm0.15$ & $9.62 \pm0.03 $  \\
20332$+$0805                 & 20   &  0.26   & $ 7967.3\pm 8.9 $&  $ 326.1\pm17.9 $& $ 0.4\pm0.1 $ &  $ 114.6\pm0.2  $ &  $10.82\pm0.04$& $11.07\pm0.12$ & $9.09 \pm 0.1 $  \\
20369$+$0150                 & 10   &  0.96   & $ 4027.5\pm13.0 $&  $ 430.7\pm26.0 $& $ 1.7\pm0.5 $ &  $  57.3\pm0.2  $ &  $10.11\pm0.06$& $10.39\pm0.13$ & $9.12 \pm0.15 $  \\
20381$+$0325                 & 10   &  0.37   & $ 8037.5\pm 4.1 $&  $ 486.7\pm 8.1 $& $ 2.4\pm0.2 $ &  $ 115.6\pm0.1  $ &  $10.72\pm0.07$& $10.97\pm0.15$ & $9.88 \pm0.03 $  \\
20415$+$1219                 & 5    &  0.49   & $ 4640.8\pm 1.3 $&  $ 361.3\pm 2.6 $& $ 9.9\pm0.2 $ &  $  66.2\pm0.1  $ &  $10.44\pm0.03$& $10.71\pm0.08$ & $10.01\pm0.01 $  \\
20417$+$1214                 & 10   &  0.42   & $ 4483.7\pm 3.4 $&  $ 364.5\pm 6.9 $& $ 3.7\pm0.2 $ &  $  63.9\pm<0.1 $ &  $10.23\pm0.05$& $10.51\pm0.12$ & $9.55 \pm0.02 $  \\
20480$+$0937                 & 5    &  0.46   & $ 4391.2\pm 2.1 $&  $ 329.2\pm 4.3 $& $ 1.7\pm0.2 $ &  $  62.6\pm0.1  $ &  $10.18\pm0.04$& $10.42\pm0.15$ & $9.20 \pm0.04 $  \\
20491$+$1846                 & 10   &  0.31   & $ 8715.1\pm 5.7 $&  $ 504.3\pm11.3 $& $ 2.3\pm0.1 $ &  $ 125.6\pm0.1  $ &  $10.92\pm0.04$& $11.18\pm0.08$ & $9.93 \pm0.03 $  \\
20550$+$1656                 & 10   &  0.29   & $10824.7\pm 1.9 $&  $ 282.0\pm 3.8 $& $ 2.1\pm0.1 $ &  $ 156.8\pm<0.1 $ &  $11.64\pm0.04$& $11.92\pm0.08$ & $10.09\pm0.02 $  \\
21052$+$0340                 & 15   &  0.27   & $ 7792.3\pm 3.6 $&  $ 518.2\pm 7.2 $& $ 1.7\pm0.1 $ &  $ 112.0\pm0.1  $ &  $10.77\pm0.05$& $11.10\pm0.08$ & $9.70 \pm0.03 $  \\
21054$+$2314$^{\dag}$        & 15   &  0.28   & $14667.6\pm 3.5 $&  $ 207.6\pm 7.1 $& $ 0.5\pm0.1 $ &  $ 214.5\pm0.1  $ &  $11.26\pm0.06$& $11.58\pm0.06$ & $9.73 \pm0.07 $  \\
21116$+$0158$^{\dag}$        & 20   &  0.21   & $ 3847.7\pm 5.3 $&  $ 519.2\pm10.7 $& $ 0.5\pm0.1 $ &  $  54.7\pm0.1  $ &  $10.28\pm0.04$& $10.54\pm0.10$ & $8.55 \pm0.07 $  \\
21271$+$0627                 & 10   &  0.25   & $ 3454.2\pm 3.3 $&  $ 262.3\pm 6.6 $& $ 1.2\pm0.1 $ &  $  49.1\pm0.1  $ &  $10.12\pm0.03$& $10.38\pm0.09$ & $8.83 \pm0.03 $  \\
21278$+$2629                 & 20   &  0.63   & $ 4837.0\pm 0.6 $&  $ 157.1\pm 1.3 $& $ 6.8\pm0.2 $ &  $  69.0\pm<0.1 $ &  $10.47\pm0.05$& $10.74\pm0.09$ & $9.88 \pm0.01 $  \\
21561$+$1148                 & 15   &  0.20   & $ 9343.9\pm 1.3 $&  $ 240.7\pm 2.6 $& $ 2.0\pm0.1 $ &  $ 134.8\pm0.1  $ &  $10.91\pm0.07$& $11.31\pm0.07$ & $9.93 \pm0.02 $  \\
21582$+$1018$^{\dag}$        & 20   &  0.22   & $ 7972.3\pm 1.4 $&  $ 295.4\pm 2.7 $& $ 1.8\pm0.1 $ &  $ 114.6\pm0.1  $ &  $10.96\pm0.04$& $11.22\pm0.08$ & $9.75 \pm0.02 $  \\
22032$+$0512                 & 20   &  0.31   & $11638.4\pm 9.2 $&  $ 357.1\pm18.5 $& $ 0.5\pm0.1 $ &  $ 168.9\pm0.2  $ &  $11.00\pm0.05$& $11.28\pm0.18$ & $9.53 \pm0.10 $  \\
22045$+$0959$^{\dag\diamond}$& 20   &  0.21   & $ 7981.3\pm 4.8 $&  $ 563.3\pm 9.7 $& $ 3.3\pm0.2 $ &  $ 114.8\pm0.3  $ &  $10.81\pm0.05$& $11.15\pm0.10$ & $10.01\pm0.02 $  \\
22171$+$2908                 & 10   &  0.45   & $ 4581.8\pm 0.9 $&  $ 442.8\pm 1.8 $& $ 8.7\pm0.2 $ &  $  65.3\pm<0.1 $ &  $10.67\pm0.08$& $10.91\pm0.10$ & $9.94 \pm0.01 $  \\
22217$+$3310                 & 10   &  0.35   & $ 6528.5\pm 0.7 $&  $ 437.4\pm 1.4 $& $ 6.3\pm0.1 $ &  $  93.5\pm<0.1 $ &  $10.55\pm0.05$& $10.81\pm0.16$ & $10.11\pm0.01 $  \\
22221$+$1748                 & 10   &  0.33   & $ 6098.5\pm 1.1 $&  $ 358.1\pm 2.2 $& $ 7.8\pm0.1 $ &  $  87.3\pm<0.1 $ &  $10.53\pm0.07$& $10.79\pm0.13$ & $10.15\pm0.01 $  \\
22347$+$3409$^{\diamond}$    & 10   &  0.54   & $  800.4\pm 0.1 $&  $ 508.1\pm 0.3 $& $148.1\pm0.3$ &  $  11.3\pm<0.1 $ &  $9.85 \pm0.06$& $10.06\pm0.08$ & $9.65 \pm<0.01$  \\
22387$+$3154                 & 10   &  0.36   & $ 7326.1\pm 6.5 $&  $ 316.9\pm13.1 $& $ 0.8\pm0.1 $ &  $ 105.2\pm0.1  $ &  $10.83\pm0.06$& $11.12\pm0.10$ & $9.32 \pm0.07 $  \\
22388$+$3359                 & 20   &  0.27   & $ 6480.6\pm 4.6 $&  $ 345.7\pm 9.2 $& $ 1.1\pm0.1 $ &  $  92.8\pm0.1  $ &  $11.08\pm0.06$& $11.33\pm0.15$ & $9.35 \pm0.04 $  \\
22395$+$2000                 & 10   &  0.39   & $ 7144.6\pm 2.8 $&  $ 426.0\pm 5.6 $& $ 3.3\pm0.2 $ &  $ 102.5\pm0.1  $ &  $10.61\pm0.05$& $11.01\pm0.16$ & $9.91 \pm0.02 $  \\
22402$+$2914                 & 20   &  0.26   & $ 7258.9\pm 1.4 $&  $ 602.5\pm 2.8 $& $ 2.0\pm0.1 $ &  $ 104.2\pm<0.1 $ &  $10.68\pm0.15$& $10.95\pm0.21$ & $9.71 \pm0.02 $  \\
22449$+$0757                 & 25   &  0.23   & $11040.4\pm20.2 $&  $ 348.2\pm40.4 $& $ 0.2\pm0.1 $ &  $ 160.0\pm0.3  $ &  $11.08\pm0.06$& $11.33\pm0.16$ & $9.08 \pm0.19 $  \\
22472$+$3439                 & 10   &  0.36   & $ 7107.0\pm 1.5 $&  $ 463.5\pm 3.0 $& $ 3.7\pm0.1 $ &  $ 102.0\pm<0.1 $ &  $10.97\pm0.05$& $11.21\pm0.08$ & $9.96 \pm0.02 $  \\
22501$+$2427                 & 20   &  0.25   & $12641.0\pm 1.7 $&  $ 220.5\pm 3.5 $& $ 0.6\pm0.1 $ &  $ 183.9\pm0.1  $ &  $11.50\pm0.13$& $11.75\pm0.17$ & $9.68 \pm0.05 $  \\
22523$+$3156$^{\dag}$        & 25   &  0.23   & $ 6373.9\pm11.3 $&  $ 724.8\pm22.7 $& $ 0.8\pm0.1 $ &  $  91.3\pm0.2  $ &  $10.52\pm0.05$& $10.79\pm0.14$ & $9.20 \pm0.05 $  \\
22575$+$1542                 &  5   &  0.58   & $ 2190.6\pm 0.3 $&  $ 293.3\pm 0.5 $& $22.8\pm0.2 $ &  $  31.0\pm<0.1 $ &  $10.14\pm0.05$& $10.39\pm0.09$ & $9.71 \pm<0.01$  \\
22586$+$0523                 & 10   &  0.34   & $ 3461.8\pm 2.3 $&  $ 297.9\pm 4.6 $& $ 2.0\pm0.1 $ &  $  49.2\pm<0.1 $ &  $9.97 \pm0.05$& $10.25\pm0.16$ & $9.06 \pm0.03 $  \\
22595$+$1541                 &  5   &  0.60   & $ 2089.6\pm 0.5 $&  $ 692.1\pm 1.0 $& $24.2\pm0.2 $ &  $  29.6\pm<0.1 $ &  $9.96 \pm0.20$& $10.20\pm0.22$ & $9.70 \pm<0.01$  \\
23007$+$2329                 & 10   &  0.29   & $ 7758.9\pm 5.8 $&  $ 215.2\pm11.6 $& $ 0.6\pm0.1 $ &  $ 111.5\pm0.1  $ &  $10.87\pm0.07$& $11.12\pm0.12$ & $9.25 \pm0.07 $  \\
23007$+$0836$^{\dag\diamond}$& 15   &  0.47   & $ 4808.3\pm 5.6 $&  $ 550.4\pm11.2 $& $ 2.3\pm0.2 $ &  $  68.6\pm0.3  $ &  $11.28\pm0.04$& $11.60\pm0.07$ & $9.66 \pm0.04 $  \\
23011$+$0046                 & 15   &  0.35   & $12666.5\pm 5.8 $&  $ 160.3\pm11.7 $& $ 0.3\pm0.1 $ &  $ 184.3\pm0.1  $ &  $11.07\pm0.06$& $11.37\pm0.16$ & $9.38 \pm0.10 $  \\
23024$+$1203$^{\diamond}$    & 10   &  0.37   & $ 2352.2\pm 0.5 $&  $ 366.3\pm 1.0 $& $20.5\pm0.2 $ &  $  33.3\pm0.1  $ &  $10.40\pm0.07$& $10.74\pm0.08$ & $9.73 \pm0.01 $  \\
23024$+$1916                 & 15   &  0.31   & $ 7453.8\pm 3.4 $&  $ 419.3\pm 6.8 $& $ 1.0\pm0.1 $ &  $ 107.0\pm0.1  $ &  $11.13\pm0.04$& $11.36\pm0.08$ & $9.43 \pm0.05 $  \\
23031$+$1856                 & 15   &  0.29   & $ 7864.1\pm10.8 $&  $ 425.2\pm21.7 $& $ 0.4\pm0.1 $ &  $ 113.1\pm0.2  $ &  $10.58\pm0.14$& $10.91\pm0.20$ & $9.08 \pm0.13 $  \\
23050$+$0359                 & 20   &  0.28   & $14258.4\pm 8.1 $&  $ 321.3\pm16.3 $& $ 0.5\pm0.1 $ &  $ 208.3\pm0.1  $ &  $11.42\pm0.04$& $11.67\pm0.16$ & $9.71 \pm0.08 $  \\
23106$+$0603$^{\diamond}$    & 10   &  0.53   & $ 3530.3\pm 0.9 $&  $  97.0\pm 1.7 $& $ 4.4\pm0.2 $ &  $  50.2\pm<0.1 $ &  $10.27\pm0.07$& $10.49\pm0.20$ & $9.42 \pm0.02 $  \\
23121$+$0415$^{\diamond}$    &  5   &  0.61   & $ 2655.6\pm 0.4 $&  $ 455.5\pm 0.8 $& $30.2\pm0.3 $ &  $  37.7\pm<0.1 $ &  $10.71\pm0.04$& $10.95\pm0.06$ & $10.01\pm<0.01$  \\
23157$+$0618                 &  5   &  0.66   & $ 4952.6\pm 0.7 $&  $ 419.2\pm 1.4 $& $13.5\pm0.3 $ &  $  70.7\pm<0.1 $ &  $10.80\pm0.04$& $11.05\pm0.14$ & $10.20\pm0.01 $  \\
23161$+$2457                 &  5   &  0.75   & $ 8114.3\pm 4.7 $&  $ 242.0\pm 9.3 $& $ 2.6\pm0.2 $ &  $ 116.7\pm0.1  $ &  $10.99\pm0.06$& $11.26\pm0.09$ & $9.92 \pm0.04 $  \\
23176$+$2356                 &  5   &  0.49   & $ 9582.6\pm 0.8 $&  $ 257.7\pm 1.6 $& $ 7.2\pm0.2 $ &  $ 138.4\pm<0.1 $ &  $10.96\pm0.05$& $11.23\pm0.13$ & $10.51\pm0.01 $  \\
23179$+$2702                 & 10   &  0.32   & $ 4272.1\pm 1.4 $&  $ 306.1\pm 2.8 $& $ 3.0\pm0.1 $ &  $  60.9\pm0.1  $ &  $10.37\pm0.04$& $10.65\pm0.08$ & $9.42 \pm0.02 $  \\
23179$+$1657                 &  5   &  0.45   & $ 1628.5\pm 0.5 $&  $ 210.8\pm 1.1 $& $12.0\pm0.1 $ &  $  23.0\pm<0.1 $ &  $9.94 \pm0.04$& $10.22\pm0.06$ & $9.18 \pm0.01 $  \\
23201$+$0805                 & 20   &  0.28   & $11502.1\pm12.7 $&  $ 122.6\pm25.4 $& $ 0.1\pm0.1 $ &  $ 166.9\pm0.2  $ &  $11.13\pm0.15$& $11.40\pm0.19$ & $8.82 \pm0.34 $  \\
23204$+$0601                 & 20   &  0.16   & $16725.6\pm 4.1 $&  $ 350.9\pm 8.2 $& $ 0.4\pm0.1 $ &  $ 245.9\pm0.1  $ &  $11.57\pm0.05$& $11.88\pm0.16$ & $9.76 \pm0.07 $  \\
23213$+$0923                 & 15   &  0.28   & $ 3580.7\pm 3.8 $&  $ 347.2\pm 7.5 $& $ 0.4\pm0.1 $ &  $  50.9\pm0.1  $ &  $10.32\pm0.05$& $10.57\pm0.11$ & $8.39 \pm0.10 $  \\
23252$+$2318                 &  5   &  0.59   & $ 3404.2\pm 0.7 $&  $ 181.0\pm 1.3 $& $ 7.0\pm0.2 $ &  $  48.4\pm<0.1 $ &  $10.26\pm0.04$& $10.50\pm0.12$ & $9.59 \pm0.01 $  \\
23254$+$0830$^{\dag\diamond}$&  5   &  0.73   & $ 8833.1\pm 2.7 $&  $ 452.1\pm 5.5 $& $ 2.8\pm0.3 $ &  $ 127.3\pm<0.1 $ &  $11.16\pm0.05$& $11.58\pm0.07$ & $10.03\pm0.04 $  \\
23256$+$2315                 &  5   &  0.58   & $ 3555.3\pm 0.5 $&  $ 277.7\pm 1.1 $& $ 8.0\pm0.2 $ &  $  50.5\pm0.1  $ &  $10.21\pm0.03$& $10.52\pm0.08$ & $9.68 \pm0.01 $  \\
23259$+$2208                 &  5   &  0.51   & $ 3488.2\pm 0.6 $&  $ 306.8\pm 1.2 $& $ 8.9\pm0.2 $ &  $  49.6\pm<0.1 $ &  $10.49\pm0.06$& $10.76\pm0.08$ & $9.71 \pm0.01 $  \\
23262$+$0314                 &  5   &  0.56   & $ 5141.8\pm 1.3 $&  $ 299.1\pm 2.7 $& $ 5.3\pm0.2 $ &  $  73.4\pm<0.1 $ &  $10.79\pm0.08$& $11.10\pm0.10$ & $9.83 \pm0.02 $  \\
23277$+$1529                 &  5   &  0.41   & $ 4219.5\pm 1.3 $&  $ 322.5\pm 2.6 $& $ 4.8\pm0.2 $ &  $  60.1\pm<0.1 $ &  $10.27\pm0.05$& $10.55\pm0.12$ & $9.61 \pm0.01 $  \\
23327$+$2913                 & 45   &  0.28   & $32145.4\pm 3.9 $&  $ 488.6\pm 7.7 $& $ 0.5\pm0.1 $ &  $ 490.1\pm0.1  $ &  $11.89\pm0.07$& $12.14\pm0.19$ & $10.45\pm0.09 $  \\
23336$+$0152                 &  5   &  0.73   & $ 2795.6\pm 0.6 $&  $ 214.6\pm 1.2 $& $13.3\pm0.2 $ &  $  39.7\pm<0.1 $ &  $10.38\pm0.06$& $10.72\pm0.08$ & $9.69 \pm0.01 $  \\
23381$+$2654                 & 10   &  0.35   & $10177.0\pm 4.7 $&  $ 302.1\pm 9.4 $& $ 1.3\pm0.1 $ &  $ 147.2\pm0.1  $ &  $10.91\pm0.05$& $11.23\pm0.17$ & $9.82 \pm0.04 $  \\
23387$+$2516                 & 10   &  0.41   & $ 9403.5\pm 2.1 $&  $ 465.0\pm 4.2 $& $ 3.5\pm0.2 $ &  $ 135.7\pm0.1  $ &  $11.04\pm0.06$& $11.31\pm0.13$ & $10.18\pm0.02 $  \\
23410$+$0228*                & 20   &  0.40   & $27335.0\pm96.0 $&  $ 400.0        $& $ 0.5       $ &  $ 412.3\pm1.6  $ &  $11.74\pm0.07$& $12.08\pm0.17$ & $< 10.30      $  \\
23414$+$0014                 & 10   &  0.42   & $ 6762.8\pm 2.5 $&  $ 436.5\pm 4.9 $& $ 2.4\pm0.2 $ &  $  97.0\pm0.1  $ &  $10.86\pm0.05$& $11.12\pm0.10$ & $9.73 \pm0.03 $  \\
23433$+$1147                 &  5   &  0.66   & $ 4262.5\pm 0.9 $&  $ 322.2\pm 1.9 $& $ 8.4\pm0.3 $ &  $  60.7\pm<0.1 $ &  $10.32\pm0.04$& $10.61\pm0.11$ & $9.86 \pm0.01 $  \\
23446$+$1519                 & 10   &  0.25   & $ 7864.3\pm 1.6 $&  $  91.0\pm 3.2 $& $ 0.5\pm0.0 $ &  $ 113.1\pm0.1  $ &  $10.87\pm0.05$& $11.22\pm0.12$ & $9.18 \pm0.04 $  \\
23456$+$2056                 & 10   &  0.35   & $ 6673.1\pm 2.7 $&  $ 457.0\pm 5.4 $& $ 2.3\pm0.1 $ &  $  95.6\pm0.1  $ &  $10.71\pm0.04$& $10.95\pm0.13$ & $9.70 \pm0.03 $  \\
23471$+$2939                 & 10   &  0.30   & $ 5268.8\pm 1.2 $&  $ 402.1\pm 2.3 $& $ 4.4\pm0.1 $ &  $  75.2\pm0.1  $ &  $10.43\pm0.06$& $10.72\pm0.11$ & $9.74 \pm0.01 $  \\
23485$+$1952                 &  5   &  0.41   & $ 4227.6\pm 1.6 $&  $ 347.7\pm 3.2 $& $ 5.1\pm0.2 $ &  $  60.2\pm<0.1 $ &  $10.77\pm0.13$& $10.99\pm0.14$ & $9.64 \pm0.01 $  \\
23488$+$1949$^{\diamond}$    &  5   &  1.37   & $ 4362.5\pm 5.3 $&  $ 815.4\pm10.6 $& $23.7\pm1.0 $ &  $  62.2\pm0.1  $ &  $11.13\pm0.05$& $11.36\pm0.07$ & $10.34\pm0.02 $  \\
23488$+$2018$^{\diamond}$    & 10   &  0.45   & $ 5392.5\pm 1.7 $&  $ 346.8\pm 3.5 $& $ 7.9\pm0.3 $ &  $  77.0\pm0.1  $ &  $11.21\pm0.04$& $11.47\pm0.06$ & $10.04\pm0.02 $  \\
23532$+$2513                 & 30   &  0.20   & $17395.3\pm 3.5 $&  $ 784.5\pm 7.0 $& $ 3.0\pm0.1 $ &  $ 256.1\pm0.1  $ &  $11.30\pm0.05$& $11.57\pm0.19$ & $10.67\pm0.02 $  \\
23560$+$1026                 &  5   &  0.25   & $ 5227.1\pm 0.7 $&  $ 302.0\pm 1.4 $& $ 4.8\pm0.1 $ &  $  74.6\pm<0.1 $ &  $10.53\pm0.05$& $10.80\pm0.11$ & $9.80 \pm0.01 $  \\
23564$+$1833                 &  5   &  0.50   & $ 5375.4\pm 1.1 $&  $ 319.6\pm 2.2 $& $ 5.2\pm0.2 $ &  $  76.8\pm<0.1 $ &  $10.41\pm0.04$& $10.70\pm0.13$ & $9.86 \pm0.02 $  \\
23568$+$2028                 &  5   &  0.49   & $ 2402.4\pm 0.5 $&  $ 109.9\pm 1.0 $& $ 4.3\pm0.1 $ &  $  34.1\pm0.1  $ &  $10.01\pm0.04$& $10.29\pm0.08$ & $9.07 \pm0.01 $  \\
23587$+$1249                 &  5   &  0.38   & $ 5317.2\pm 4.0 $&  $ 357.5\pm 7.9 $& $ 3.3\pm0.2 $ &  $  75.9\pm0.1  $ &  $10.33\pm0.05$& $10.60\pm0.15$ & $9.65 \pm0.02 $  \\
23591$+$2312$^{\diamond}$    &  5   &  0.51   & $ 4366.1\pm 1.8 $&  $ 627.5\pm 3.5 $& $ 7.3\pm0.2 $ &  $  62.2\pm<0.1 $ &  $10.66\pm0.06$& $10.91\pm0.11$ & $9.82 \pm0.01 $  \\
23597$+$1241                 & 10   &  0.33   & $ 5487.7\pm 1.4 $&  $ 417.7\pm 2.9 $& $ 3.9\pm0.1 $ &  $  78.4\pm<0.1 $ &  $10.57\pm0.05$& $10.86\pm0.09$ & $9.75 \pm0.02 $  \\
\enddata
\tablenotetext{\diamond}{Sources observed in DPS mode.}
\tablenotetext{*}{HI 21 cm non-detections.}
\tablenotetext{\dag}{Sources that exhibit both emission and absorption features, but with parameters derived from the
emission part of the spectra. The flux density integrals and neutral hydrogen masses should be considered lower
limits since some of the emission might be masked by absorption features.}
\end{deluxetable}

\begin{deluxetable}{lccccccccc}
\tabletypesize{\footnotesize}
\rotate
\tablenum{3}
\tablecolumns{10}
\tablewidth{0pc}
\tablecaption{Parameters of Galaxies with \HI Absorption}
\tablehead{
\colhead{IRAS Name}
& \colhead{$D_{\rm L}$\tablenotemark{a}}
& \colhead{log $\frac{L_{\rm FIR}}{\rm L_ {\odot}}$}
& \colhead{log $\frac{L_{\rm IR}}{\rm L_ {\odot}}$}
& \colhead{$\Delta t$}
& \colhead{rms}
& \colhead{$V_{\rm HI-Peak}$}
& \colhead{$\Delta{V_{50}}$}
& \colhead{$\tau_{\rm max} \times 10^{-2}$}
& \colhead{$N_{\rm HI}/T_{\rm s} \times 10^{18}$} \\
\colhead{}
& \colhead{(Mpc)}
& \colhead{}
& \colhead{}
& \colhead{(min)}
& \colhead{(mJy)}
& \colhead{(km $\rm s^{-1}$)}
& \colhead{(km $\rm s^{-1}$)}
& \colhead{}
& \colhead{(cm$^{-2}$ K$^{-1}$)} \\
\colhead{(1)}
& \colhead{(2)}
& \colhead{(3)}
& \colhead{(4)}
& \colhead{(5)}
& \colhead{(6)}
& \colhead{(7)}
& \colhead{(8)}
& \colhead{(9)}
& \colhead{(10)}
}
\startdata
21054$+$2314$^{\dag}              $  &$ 214.5\pm 0.1$ &$ 11.26\pm0.06 $ &$ 11.58\pm0.06 $ & 15 &0.28&$14893\pm9  $ &$ 144\pm24  $ &$ 7.1 \pm0.2 $  &$ 12.6\pm0.4 $  \\
21116$+$0158$^{\dag}              $  &$ 54.7 \pm 0.1$ &$ 10.28\pm0.04 $ &$ 10.54\pm0.10 $ & 20 &0.21&$4079 \pm11 $ &$  94\pm9   $ &$ 17.9\pm1.8 $  &$ 19.5\pm0.7 $  \\
21442$+$0007                         &$ 330.9\pm 0.4$ &$ 11.59\pm0.05 $ &$ 11.84\pm0.17 $ & 25 &0.31&$22241\pm6  $ &$ 101\pm21  $ &$ 11.0\pm2.2 $  &$ 15.3\pm0.9 $  \\
21582$+$1018$^{\dag}              $  &$ 114.6\pm 0.1$ &$ 10.96\pm0.04 $ &$ 11.22\pm0.08 $ & 20 &0.22&$8155 \pm11 $ &$ 141\pm40  $ &$ 4.0 \pm0.5 $  &$  4.1\pm0.2 $  \\
22045$+$0959$^{\dag\diamond}      $  &$ 114.8\pm 0.3$ &$ 10.81\pm0.05 $ &$ 11.15\pm0.10 $ & 20 &0.21&$7982 \pm3  $ &$  62\pm20  $ &$ 6.1 \pm0.6 $  &$  7.9\pm0.3 $  \\
22523$+$3156$^{\dag}              $  &$  91.3\pm 0.2$ &$ 10.52\pm0.05 $ &$ 10.79\pm0.14 $ & 25 &0.23&$6260 \pm8  $ &$ 185\pm20  $ &$ 33.8\pm1.3 $  &$ 44.0\pm0.7 $  \\
23007$+$0836$^{\dag\diamond}      $  &$  68.6\pm 0.3$ &$ 11.28\pm0.04 $ &$ 11.60\pm0.07 $ & 15 &0.47&$4916 \pm16 $ &$  75\pm5   $ &$ 7.2 \pm0.4 $  &$  9.5\pm0.2 $  \\
23254$+$0830$^{\dag\diamond}      $  &$ 127.3\pm<0.1$ &$ 11.16\pm0.05 $ &$ 11.58\pm0.07 $ & 5  &0.73&$8571 \pm8  $ &$  63\pm4   $ &$ 7.6 \pm0.3 $  &$ 17.9\pm0.2 $  \\
23365$+$3604                         &$ 286.1\pm 0.2$ &$ 11.94\pm0.05 $ &$ 12.17\pm0.15 $ & 25 &0.30&$19380\pm6  $ &$  91\pm8   $ &$ 9.2 \pm0.8 $  &$ 23.8\pm0.5 $  \\
23594$+$3622$^{\diamond {\rm b}}  $  &$ 139.0\pm 0.6$ &$ 11.11\pm0.06 $ &$ 11.34\pm0.15 $ & 15 &0.62&$9670 \pm6  $ &$ 162\pm58  $ &$ 6.1 \pm0.8 $  &$ 16.1\pm0.5 $  \\
\enddata
\tablenotetext{\diamond}{Sources observed in DPS mode.}
\tablenotetext{\dag}{Sources that exhibit both emission and absorption features, but with \HI parameters derived from the absorption part of the spectra.}
\tablenotetext{a}{Derived using the \HI emission velocity (Table~2) for sources that exhibit both \HI emission and absorption,
and optical redshifts (Table~1) for sources that exhibit \HI absorption only.}
\tablenotetext{b}{The \HI spectrum for this source (Figure~2) shows another weak absorption-like features at 9350~km~s$^{-1}$.
However, it only has a $2.5\sigma$ significance.}
\end{deluxetable}

\begin{deluxetable}{lccccccccccccccc}
\tabletypesize{\scriptsize}
\center
\rotate
\tablenum{4}
\tablecolumns{16}
\tablewidth{0pc}
\tablecaption{Parameters of Galaxies with OH Megamaser Emission}
\tablehead{
\colhead{}
& \colhead{}
& \colhead{}
& \multicolumn{3}{c}{1667 MHz}
& \colhead{}
& \multicolumn{3}{c}{1665 MHz}
& \colhead{}
& \colhead{}
& \colhead{} \\
\cline{4-7}  \cline{9-11} \\
\colhead{IRAS Name}
& \colhead{$\Delta t$}
& \colhead{rms}
& \colhead{$V$}
& \colhead{$F_{\rm peak}$}
& \colhead{$\Delta{V_{50}}$}
& \colhead{$\int{Sdv}$}
& \colhead{}
& \colhead{$F_{\rm peak}$}
& \colhead{$\Delta{V_{50}}$}
& \colhead{$\int{Sdv}$}
& \colhead{$R_{\rm H}$}
& \colhead{log $\frac{L_{\rm OH}^{\rm pred}}{\rm L_{\odot}}$}
& \colhead{log $\frac{L_{\rm OH}}{\rm L_{\odot}}$}
& \colhead{rms$_{\rm 1612}$}
& \colhead{rms$_{\rm 1720}$} \\
\colhead{}
& \colhead{(min)}
& \colhead{(mJy)}
& \colhead{(km $\rm s^{-1}$)}
& \colhead{(mJy)}
& \colhead{(km $\rm s^{-1}$)}
& \colhead{(Jy km $\rm s^{-1}$)}
& \colhead{}
& \colhead{(mJy)}
& \colhead{(km $\rm s^{-1}$)}
& \colhead{(Jy km $\rm s^{-1}$)}
& \colhead{}
& \colhead{}
& \colhead{}
& \colhead{(mJy)}
& \colhead{(mJy)} \\
\colhead{(1)}
& \colhead{(2)}
& \colhead{(3)}
& \colhead{(4)}
& \colhead{(5)}
& \colhead{(6)}
& \colhead{(7)}
& \colhead{}
& \colhead{(8)}
& \colhead{(9)}
& \colhead{(10)}
& \colhead{(11)}
& \colhead{(12)}
& \colhead{(13)}
& \colhead{(14)}
& \colhead{(15)}}
\startdata
23050+0359 & 20 &  0.19  & $14239\pm9.6$ & $ 4.0\pm0.3$ & $197\pm13$ & $ 0.50\pm0.01$ & & $2.0\pm0.3$ & $200\pm50$ & $0.24\pm0.01$ &$2.08\pm0.10$&   1.74  & 1.69 & --   & --  \\
23327+2913 & 45 &  0.19  & $32285\pm5.4$ & $ 1.4\pm0.2$ & $204\pm34$ & $ 0.30\pm0.01$ & &   --        &    --      &   --          &  --         &   2.39  & 2.04 & 0.23 & --  \\
23365+3604 & 25 &  0.28  & $19222\pm7.4$ & $ 7.1\pm0.3$ & $256\pm56$ & $ 1.90\pm0.01$ & & $2.4\pm0.3$ & $ 172\pm70$ &$0.50\pm0.01$ &$3.80\pm0.08$&   2.46  & 2.46 & 0.21 & --  \\
\enddata
\end{deluxetable}

\begin{deluxetable}{lcccccccccccccc}
\tabletypesize{\scriptsize}
\rotate
\tablenum{5}
\tablecolumns{15}
\tablewidth{0pc}
\tablecaption{Parameters of Galaxies with OH Absorption}
\tablehead{
\colhead{}
& \colhead{}
& \colhead{}
& \multicolumn{5}{c}{1667 MHz}
& \colhead{}
& \multicolumn{3}{c}{1665 MHz}
& \colhead{} \\
\cline{4-8}  \cline{10-12} \\
\colhead{IRAS Name}
& \colhead{$\Delta t$}
& \colhead{rms}
& \colhead{$V$}
& \colhead{$\Delta{V_{50}}$}
& \colhead{$\tau \times 10^{-2}$}
& \colhead{$\int{\tau dv}$}
& \colhead{$N_{\rm OH}/T_{\rm ex}\times 10^{14}$}
& \colhead{}
& \colhead{$\Delta{V_{50}}$}
& \colhead{$\tau \times 10^{-2}$}
& \colhead{$\int{\tau dv}$}
& \colhead{$R_{\rm H}$}
& \colhead{rms$_{\rm 1612}$}
& \colhead{rms$_{\rm 1720}$} \\
\colhead{}
& \colhead{(min)}
& \colhead{(mJy)}
& \colhead{(km $\rm s^{-1}$)}
& \colhead{(km $\rm s^{-1}$)}
& \colhead{}
& \colhead{(km $\rm s^{-1}$)}
& \colhead{($\rm cm^{-2} K^{-1}$)}
& \colhead{}
& \colhead{(km $\rm s^{-1}$)}
& \colhead{}
& \colhead{(km $\rm s^{-1}$)}
& \colhead{}
& \colhead{(mJy)}
& \colhead{(mJy)} \\
\colhead{(1)}
& \colhead{(2)}
& \colhead{(3)}
& \colhead{(4)}
& \colhead{(5)}
& \colhead{(6)}
& \colhead{(7)}
& \colhead{(8)}
& \colhead{}
& \colhead{(9)}
& \colhead{(10)}
& \colhead{(11)}
& \colhead{(12)}
& \colhead{(13)}
& \colhead{(14)}}
\startdata
21054+2314   & 40 & 0.68  & $14792\pm27$ & $133\pm14$ & $6.5\pm1.0$  & $7.4\pm0.3$ & $17.4\pm0.7$ & & --        &  --           &  --         &  --          & --   & 0.31  \\
22523+3156   & 25 & 0.22  & $ 6250\pm2$  & $110\pm24$ & $16.9\pm0.6$ & $13.0\pm0.1$& $30.6\pm0.3$ & & $76\pm3$  &  $11.0\pm0.5$ & $6.9\pm0.1$ &$1.88\pm0.03 $& --   & 0.18  \\
23007+0836   & 15 & 0.85  & $ 4947\pm16$ & $200\pm100$& $2.1\pm0.3$  & $4.4\pm0.1$ & $10.3\pm0.2$ & & $98\pm51$ &  $1.9\pm0.3$  & $2.3\pm0.1$ &$1.91\pm0.09 $& --   & 0.18  \\
23121+0415   &  5 & 0.72  & $ 2650\pm9$  & $218\pm68$ & $2.3\pm0.5$  & $2.4\pm0.1$ & $ 5.6\pm0.3$ & & $108\pm65$&  $2.4\pm0.5$  & $2.7\pm0.1$ &$0.89\pm0.05 $& 0.69 & --    \\
\enddata
\end{deluxetable}

\begin{deluxetable}{lcrrcc}
\tablenum{6}
\tablecolumns{6}
\tablewidth{0pc}
\tablecaption{Parameters of Galaxies with OH Nondetections}
\tablehead{
\colhead{IRAS Name}
& \colhead{rms$_{1667,1665}$}
& \colhead{log $\frac{L_{\rm OH}^{\rm pred}}{\rm L_{\odot}}$}
& \colhead{log $\frac{L_{\rm OH}^{\rm max}}{\rm L_{\odot}}$}
& \colhead{rms$_{\rm 1612}$}
& \colhead{rms$_{\rm 1720}$} \\
\colhead{}
& \colhead{(mJy)}
& \colhead{}
& \colhead{}
& \colhead{(mJy)}
& \colhead{(mJy)} \\
\colhead{(1)}
& \colhead{(2)}
& \colhead{(3)}
& \colhead{(4)}
& \colhead{(5)}
& \colhead{(6)}}
\startdata
20082+0058   &  -      &   0.71   &   -      &    -     &       -    \\
20093+0536   &  0.31   &   0.64   & $-0.17$  &    0.24  &       -    \\
20178-0052   &  0.91   &   0.48   &   0.34   &    -     &       -    \\
20198+0159   &  -      &   1.31   &   -      &    -     &       0.30 \\
20210+1121   &  -      &   1.78   &   -      &    -     &       -    \\
20230+1024   &  -      &   0.66   &   -      &    -     &       0.24 \\
20332+0805   &  -      &   0.91   &   -      &    0.38  &       0.42 \\
20369+0150   &  1.00   & $-0.07$  &   0.10   &    0.80  &       -    \\
20381+0325   &  -      &   0.77   &   -      &    -     &       0.25 \\
20415+1219   &  0.39   &   0.39   & $-0.18$  &    0.40  &       0.52 \\
20417+1214   &  0.32   &   0.09   & $-0.30$  &    0.31  &       -    \\
20480+0937   &  0.44   &   0.03   & $-0.17$  &    0.45  &       0.46 \\
20491+1846   &  -      &   1.05   &   -      &    -     &       0.27 \\
20550+1656   &  0.24   &   2.04   &   0.35   &    -     &       0.23 \\
21052+0340   &  -      &   0.84   &   -      &    -     &       0.23 \\
21116+0158   &  -      &   0.17   &   -      &    0.22  &       0.14 \\
21271+0627   &  0.28   & $-0.06$  & $-0.58$  &    0.23  &       -    \\
21278+2629   &  0.46   &   0.43   & $-0.07$  &    0.23  &       0.49 \\
21442+0007   &  -      &   1.98   &   -      &    0.34  &       -    \\
21561+1148   &  0.30   &   1.03   &   0.31   &    0.26  &       0.26 \\
21582+1018   &  -      &   1.10   &   -      &    -     &       0.28 \\
22032+0512   &  -      &   1.17   &   -      &    -     &       0.23 \\
22045+0959   &  -      &   0.90   &   -      &    -     &       0.79 \\
22171+2908   &  0.40   &   0.71   & $-0.18$  &    0.41  &       -    \\
22217+3310   &  -      &   0.53   &   -      &    -     &       0.33 \\
22221+1748   &  0.36   &   0.51   &   0.03   &    -     &       0.53 \\
22347+3409   &  0.95   & $-0.42$  & $-1.33$  &    -     &       -    \\
22387+3154   &  -      &   0.93   &   -      &    -     &       0.30 \\
22388+3359   &  0.35   &   1.27   &   0.06   &    -     &       0.24 \\
22395+2000   &  -      &   0.62   &   -      &    -     &       0.32 \\
22402+2914   &  -      &   0.72   &   -      &    -     &       0.15 \\
22449+0757   &  -      &   1.27   &   -      &    -     &       0.16 \\
22472+3439   &  -      &   1.12   &   -      &    -     &       0.33 \\
22501+2427   &  -      &   1.85   &   -      &    -     &       0.17 \\
22575+1542   &  0.69   & $-0.03$  & $-0.59$  &    -     &       -    \\
22586+0523   &  0.27   & $-0.27$  & $-0.60$  &    0.39  &       0.42 \\
22595+1541   &  0.51   & $-0.27$  & $-0.76$  &    -     &       0.79 \\
23007+2329   &  -      &   0.97   &   -      &    -     &       0.29 \\
23011+0046   &  -      &   1.26   &   -      &    -     &       0.21 \\
23024+1203   &  0.36   &   0.33   & $-0.81$  &    -     &       -    \\
23024+1916   &  -      &   1.34   &   -      &    -     &       0.29 \\
23031+1856   &  -      &   0.58   &   -      &    -     &       0.21 \\
23106+0603   &  -      &   0.15   &   -      &    -     &       -    \\
23157+0618   &  0.42   &   0.88   & $-0.09$  &    -     &       0.46 \\
23161+2457   &  -      &   1.14   &   -      &    0.87  &       0.59 \\
23176+2356   &  0.42   &   1.10   &   0.48   &    0.51  &       0.39 \\
23179+2702   &  0.37   &   0.29   & $-0.27$  &    0.40  &       -    \\
23179+1657   &  0.34   & $-0.30$  & $-1.16$  &    -     &       0.44 \\
23201+0805   &  -      &   1.34   &   -      &    -     &       0.18 \\
23204+0601   &  -      &   1.94   &   -      &    -     &       -    \\
23213+0923   &  0.25   &   0.22   & $-0.60$  &    0.32  &       -    \\
23252+2318   &  0.36   &   0.14   & $-0.49$  &    0.64  &       -    \\
23254+0830   &  -      &   1.38   &   -      &    -     &       0.58 \\
23256+2315   &  0.44   &   0.07   & $-0.36$  &    0.39  &       -    \\
23259+2208   &  0.49   &   0.46   & $-0.33$  &    0.54  &       0.53 \\
23262+0314   &  0.46   &   0.87   & $-0.02$  &    0.41  &       -    \\
23277+1529   &  -      &   0.15   &   -      &    0.37  &       0.37 \\
23336+0152   &  0.61   &   0.30   & $-0.43$  &    0.67  &       -    \\
23381+2654   &  -      &   1.04   &   -      &    0.21  &       0.18 \\
23387+2516   &  -      &   1.22   &   -      &    -     &       0.28 \\
23410+0228   &  -      &   2.18   &   -      &    0.24  &       -    \\
23414+0014   &  -      &   0.96   &   -      &    -     &       0.33 \\
23433+1147   &  0.45   &   0.22   & $-0.19$  &    0.50  &       -    \\
23446+1519   &  -      &   0.98   &   -      &    -     &       0.34 \\
23456+2056   &  -      &   0.76   &   -      &    -     &       0.26 \\
23471+2939   &  0.59   &   0.33   &   0.08   &    0.29  &       0.39 \\
23485+1952   &  -      &   0.84   &   -      &    0.38  &       -    \\
23488+1949   &  -      &   1.34   &   -      &    1.31  &       -    \\
23488+2018   &  -      &   1.45   &   -      &    0.46  &       0.42 \\
23532+2513   &  -      &   1.58   &   -      &    -     &       -    \\
23560+1026   &  -      &   0.51   &   -      &    0.26  &       0.44 \\
23564+1833   &  -      &   0.35   &   -      &    0.44  &       0.52 \\
23568+2028   &  0.45   & $-0.21$  & $-0.69$  &    -     &       -    \\
23587+1249   &  -      &   0.23   &   -      &    0.34  &       0.52 \\
23591+2312   &  -      &   0.69   &   -      &    0.45  &       -    \\
23594+3622   &  -      &   1.32   &   -      &    1.35  &       1.33 \\
23597+1241   &  0.32   &   0.57   & $-0.12$  &    0.29  &       0.35 \\
\enddata
\end{deluxetable}

\clearpage

\begin{deluxetable}{lcccc}
\tabletypesize{\footnotesize}
\tablenum{7}
\tablecolumns{5}
\tablewidth{0pc}
\tablecaption{The Mean and Median \HI mass values within different $L_{\rm IR}$ ranges}
\tablehead{
\colhead{IR Luminosity ($L_{\odot}$)}
& \colhead{$L_{\rm IR} \leq 10^{10.49}$}
& \colhead{$10^{10.50}\leq L_{\rm IR}\leq 10^{10.99}$}
& \colhead{$10^{11.00}\leq L_{\rm IR} \leq 10^{11.49}$}
& \colhead{$L_{\rm IR}\geq 10^{11.50}$}}
\startdata
Mean \HI mass    &  $10^{9.29 \pm 0.10}$  & $10^{9.67 \pm 0.07}$  &  $10^{9.69 \pm 0.08}$  &  $10^{9.99 \pm 0.10}$    \\
Median \HI mass  &  $10^{9.19}$           & $10^{9.72}$           &  $10^{9.78}$           &  $10^{9.86}$             \\
\enddata
\end{deluxetable}

\begin{deluxetable}{lcccc}
\tabletypesize{\footnotesize}
\tablenum{8a}
\tablecolumns{5}
\tablewidth{0pc}
\tablecaption{The occurrence of \HI 21~cm absorption in galaxies within different $L_{\rm IR}$ ranges}
\tablehead{
\colhead{IR Luminosity ($L_{\odot}$)}
& \colhead{$L_{\rm IR} \leq 10^{10.49}$}
& \colhead{$10^{10.50}\leq L_{\rm IR}\leq 10^{10.99}$}
& \colhead{$10^{11.00}\leq L_{\rm IR} \leq 10^{11.49}$}
& \colhead{$L_{\rm IR}\geq 10^{11.50}$}}
\startdata
No. (abs.~or abs.+emis.) &  0    &   2   &   3 &  5    \\
Total No.~of galaxies &  10   &  32   &   29  &  13    \\
\% with Absorption &  0\%  &   6.25\%   &   10.3\% & 38.5\% \\
\enddata
\end{deluxetable}

\begin{deluxetable}{lcccc}
\tabletypesize{\footnotesize}
\tablenum{8b}
\tablecolumns{5}
\tablewidth{0pc}
\tablecaption{NVSS flux density distribution of the galaxies with \HI
absorption within different $L_{\rm IR}$ ranges}
\tablehead{
\colhead{IR Luminosity ($L_{\odot}$)}
& \colhead{{$L_{\rm IR} \leq 10^{10.49}$}}
&\colhead{{$10^{10.50}\leq L_{\rm IR}\leq 10^{10.99}$}}
& \colhead{{$10^{11.00}\leq L_{\rm IR} \leq 10^{11.49}$}}
& \colhead{$L_{\rm IR}\geq 10^{11.50}$}}
\startdata
Median NVSS Flux (mJy) &  20.40    &   24.15    &   31.5 &  15.8    \\
Mean NVSS Flux (mJy)  & $48.33\pm 20.27$   &  $37.52\pm6.26$   &  $36.6\pm5.92$  &  $47.33\pm19.43$\\
Mean Flux with abs (mJy)   &  -  &   28.9   &  82.33 & 90.22 \\
\enddata
\end{deluxetable}

\begin{deluxetable}{lcccc}
\tabletypesize{\footnotesize}
\tablenum{9}
\tablecolumns{5}
\tablewidth{0pc}
\tablecaption{The occurrence of OH 18~cm main line (absorption or emission) in galaxies within different $L_{\rm IR}$ ranges}
\tablehead{
\colhead{IR Luminosity ($L_{\odot}$)}
& \colhead{$L_{\rm IR} \leq 10^{10.49}$}
& \colhead{$10^{10.50}\leq L_{\rm IR}\leq 10^{10.99}$}
& \colhead{$10^{11.00}\leq L_{\rm IR} \leq 10^{11.49}$}
& \colhead{$L_{\rm IR}\geq 10^{11.50}$}}
\startdata
No. (abs.~or emis.) &  0    &   2   &   0 &  5    \\
Total No.~of galaxies &  9   &  19  &   5  &  6    \\
\% OH Detections &  0\%  &   10.5\%   &   0\% & 83.3\% \\
\enddata
\end{deluxetable}


\clearpage
\begin{thebibliography}{}
\bibitem[Baan et al.(1987)]{BGSM87}Baan, W. A., van Gorkom, J. H., Schmelz, J. T, Mirabel, I. F. 1987, \apj, 313, 102
\bibitem[Baan(1989)]{B89}Baan, W. A. 1989, \apj, 338, 804
\bibitem[Condon(1992)]{CON92}Condon, J. J. 1992, \araa, 30, 575
\bibitem[Condon et al.(1998)]{CON98}Condon, J. J., Cotton, W. D., Greisen, E. W., Yin, Q. F., Perley, R. A., Taylor, G. B., Broderick, J. J. 1998, \aj, 115, 1693
\bibitem[Darling \& Giovanelli(2000)]{DG00}Darling, J., \& Giovanelli, R. 2000, \aj, 119, 3003
\bibitem[Darling \& Giovanelli(2001)]{DG01}Darling, J., \& Giovanelli, R. 2001, \aj, 121, 1278
\bibitem[Darling \& Giovanelli(2002)]{DG02}Darling, J., \& Giovanelli, R. 2002, \aj, 124, 100
\bibitem[Dinh-V-Trung et al.(2001)]{DVT01}Dinh-V-Trung, Lo, K. Y., Kim, D.-C., Gao, Yu, Gruendl, R. A. 2001, \apj, 556, 141
\bibitem[Falco et al.(1999)]{FAL99}Falco, E. E., Kurtz, M. J., Geller, M. J., et al. 1999, \pasp, 111, 438
\bibitem[Ghosh \& Salter(2002)]{ghos02} Ghosh, T., \& Salter, C.J. 2002,
in ASP Conf. Ser. 278, Single-Dish Radio Astronomy: Techniques and
Applications, ed.  S. Stanimirovic, D. Altschuler, P. Goldsmith, \&
C. Salter, (San Francisco: ASP), 521
\bibitem[Helou, Soifer \& Rowan-Robinson(1985)]{HSR85}Helou, G., Soifer, B. T.,
\& Rowan-Robinson, M. 1985, \apjl, 298, L7
\bibitem[Kandalian(1996)]{KAN96}Kandalian, R. A. 1996, Astrophysics, 39, 237
\bibitem[Mirabel(1982)]{MIR82}Mirabel, I. F. 1982, \apj, 260, 75
\bibitem[Momjian et al.(2003)]{MOM03}Momjian, E., Romney, J. D., Carilli, C. L., Troland, T. H., 2003, \apj, 597, 809
\bibitem[Paturel et al.(2003)]{PAT03}Paturel, G., Petit, C., Prugniel, Ph., Theureau, G., Rousseau, J., Brouty, M., Dubois, P., Cambresy, L. 2003, \aap, 412, 45
\bibitem[Roberts(1975)]{Rob75}Roberts, M. S. 1975, in Galaxies and the Universe, ed. A. Sandage, M. Sandage, \& J. Kristian (Chicago : Univ. Chicago Press), 309
\bibitem[Rohlfs(1986)]{Roh86}Rohlfs, K., 1986, Tools of Radio Astronomy, (Springer-Verlag)
\bibitem[Sanders \& Mirabel(1996)]{SM96}Sanders, D. B., \& Mirabel, I. F. 1996, \araa, 34, 749
\bibitem[Soifer \& Neugebauer(1991)]{SN91}Soifer, B. T., \& Neugebauer, G. 1991, \aj, 101, 354
\bibitem[Soifer et al.(1987)]{SOI87}Soifer, B. T., Sanders, D. B., Madore, B. F., Neugebauer, G., Danielson, G. E., Elias, J. H., Lonsdale, Carol J., Rice, W. L. 1987, \apj, 320, 238
\bibitem[Strauss et al.(1992)]{STR92}Strauss, M. A., Huchra, J. P., Davis, M., Yahil, A., Fisher, K. B., Tonry, J. L. 1992, \apjs, 83, 29
\bibitem[Turner(1973)]{Tur73}Turner, B. E. 1973, \apj, 186, 357
\bibitem[Vorontsov-Velyaminov \& Arkhipova(1963)]{VVA93}Vorontsov-Velyaminov, B. A., \& Arkhipova, V. P. 1963, Morphological Catalogue of Galaxies, Part III (Trudy Astr. Inst. Sternberg 33), 68
\bibitem[Wasilewski(1981)]{WAS81}Wasilewski, A. J. 1981, PASP, 93, 560
\bibitem[Wong \& Blitz(2002)]{WB02}Wong, T., \& Blitz, L. 2002, \apj, 569, 157
\bibitem[Wunderlich \& Klein(1988)]{WK88}Wunderlich, E., \&
Klein, U. 1988, \aap, 206, 47
\bibitem[Yun, Reddy \& Condon(2001)]{YRC01}Yun, M. S., Reddy, N. A., \& Condon, J. J. 2001, \apj, 554, 803
\end{thebibliography}
\end{document}